\documentclass[11pt,a4paper]{article}

\usepackage[T1]{fontenc}
\usepackage[utf8]{inputenc}
\usepackage{lmodern}
\usepackage[a4paper,top=26mm,bottom=23mm,left=26mm,right=26mm]{geometry}
\usepackage{graphicx}
\usepackage{placeins}
\usepackage{float}
\usepackage{caption}
\usepackage{multirow}
\usepackage{amsmath,amssymb,amsfonts,amsthm}
\usepackage{mathrsfs}
\usepackage[title]{appendix}
\usepackage{xcolor}
\usepackage{colortbl}
\usepackage{textcomp}
\usepackage{manyfoot}
\usepackage{booktabs}
\IfFileExists{bbm.sty}{\usepackage{bbm}}{\newcommand{\mathbbm}[1]{\mathrm{##1}}}
\usepackage{mathtools}
\usepackage{tikz}
\usepackage{tikz-cd}
\usetikzlibrary{patterns,arrows,decorations.pathmorphing}
\usepackage{etoolbox}
\usepackage[numbers,sort&compress]{natbib}
\usepackage[hidelinks]{hyperref}
\newcommand{\im}{\mathbbm{i}}
  \theoremstyle{definition}
\newtheorem{theorem}{Theorem}[section]
\newtheorem{proposition}[theorem]{Proposition}

\newtheorem{corollary}[theorem]{Corollary}
\newtheorem{definition}{Definition}[section]

\newtheorem{remark}{Remark}[section]
\AtBeginEnvironment{theorem}{\normalfont}
\AtBeginEnvironment{proposition}{\normalfont}
\AtBeginEnvironment{lemma}{\normalfont}
\AtBeginEnvironment{corollary}{\normalfont}
\AtBeginEnvironment{definition}{\normalfont}
\AtBeginEnvironment{example}{\normalfont}
\AtBeginEnvironment{remark}{\normalfont}
\title{Propagaton Dynamics:\\A Geometric Theory of Relational Field Motion}

\author{Said Mikki\\[3pt]\small ZJU--UIUC Institute, Zhejiang University\\\small International Campus, Haining, Zhejiang, China\\\small\texttt{said.m.mikki@gmail.com}}
\date{}
\begin{document}
\maketitle
\begin{abstract}
We develop a geometric theory of relational field motion based on the generation of propagating waves relative to non-propagating waves within a source--response field system. Using the Weyl representation of source-generated scalar monochromatic fields, we construct a propagation potential, defined as the negative ratio of the evanescent to the propagating squared amplitude, on the six-dimensional configuration manifold $\mathcal M=\mathbb R^3\times\mathrm{SO}(3)$, incorporating spatial position and a local spectral frame. A Riemannian metric and a unified dynamical law determine the evolution of a point-like configuration, termed a \emph{propagaton}. The massless branch defines an ascending gradient flow on the configuration manifold, while the massive branch incorporates inertia and admits, in the conservative case, the Lagrangian and Hamiltonian formulations of geometric mechanics. Analytical regularity results establish local well-posedness on the regular configuration domain under suitable metric hypotheses for a broad class of source distributions, including H\"older-continuous sources of compact support and, in exterior observation regions, general bounded sources of compact support. We establish joint Euclidean covariance and identify an intrinsic $\mathrm{SO}(2)$ symmetry of the spectral partition. For axially invariant metrics, Noether's theorem gives a conserved charge in the conservative massive branch, which relaxes exponentially in the inertial dynamics with damping. The associated quotient geometry permits a five-dimensional reduction of the massless and zero-axial-momentum inertial dynamics. Analytical and numerical investigations reveal nonradial propagaton motion within a spherically symmetric radiation field and establish the contrasting dynamical significance of evanescent-free configurations for massless and massive propagatons. The theory opens a research direction in which propagatons may constitute elementary building blocks of motion arising from fundamental relations between fields.

\end{abstract}
\section{Introduction}
\label{sec:introduction}

The introduction of the field concept into physics brought about a profound change in the way physical interactions are understood. With the development of electromagnetic theory, the field came to acquire a physical significance of its own, and the description of interactions between material bodies became inseparable from the study of processes taking place in the space surrounding them~\cite{berkson2014fields,Siegel1992Innovation-in-maxwell}. The problem of motion may also be considered from this perspective. In mechanics, motion is ordinarily described in terms of the changing position of a localized object, while relational approaches to dynamics emphasize the physical systems with respect to which such changes are defined~\cite{barbour2001the_discovery_of_dynamics,barbour2000the_end_of_time}. If fields are taken as the fundamental physical entities, one may ask how the notion of local motion can arise from the relations between fields themselves. In particular, is it possible to extract from the spatial structure of a field the information needed to formulate a law of motion? The aim of this paper is to explore this question by developing a new theory of motion associated with the generation and production of propagating waves relative to non-propagating waves.

A natural starting point for such a theory is the relation between a source field $J$ and the radiation field $\varphi$ it generates. The source provides a physical reference for examining the spatial structure of the radiation field. The two fields are distinguished here by their roles in the governing wave equation, and the description of motion will be developed relative to this particular source--response relation. A useful observation concerning the nature of wave propagation is that, whereas the rigid translation of a field profile can accommodate arbitrary transverse structure, a pure longitudinal monochromatic mode is necessarily a plane wave (Proposition~\ref{thm:monochromatic_pure_mode}). The radiation field produced by a localized source generally contains a spectrum of propagating and evanescent components. By resolving this superposition relative to a selected direction, one can examine the extent to which the field at a given point exhibits the characteristics of propagation along that direction. This suggests that the spectral structure of a source-generated field might provide a means of defining local motion. The mathematical basis for this approach is the Weyl representation of wave fields~\cite{Weyl1919Propagation}, which has found extensive applications in optics and electromagnetic theory~\cite{clemmow1996the,hansen1999plane_wave,rfdFelsenMarcuvitz1994}. Relative to a selected longitudinal axis, the representation separates the angular spectrum into propagating modes with real longitudinal wavenumbers and evanescent modes with imaginary longitudinal wavenumbers. The latter decay along the longitudinal direction and can have oscillatory transverse structure. The contribution of evanescent modes to the electromagnetic Green tensor, including their directional behavior at large distances from the source, has been investigated in earlier studies~\cite{ShchegrovCarney1999,Arnoldus_Foley_2002}. An important feature of the spectral representation for the purposes of the present work is the role played by the choice of the longitudinal axis. For a fixed source and observation point, rotating this axis changes the classification of spectral contributions and can therefore change the values of the propagating and evanescent partial fields, whose sum reconstructs the same total response. We examine this dependence explicitly in Section~\ref{sec:rotated_weyl_field_decomposition} and Appendix~\ref{app:rfd_point_source_rotation}.

The theory developed here grew out of the author's earlier investigations of the spatial structure of electromagnetic radiation, published between 2011 and 2015. A spectral theory of the electromagnetic near field was developed in~\cite{mikki_nf1,mikki_nf2}, where the separation into propagating and non-propagating contributions provided a means of studying near-field streamlines, energy localization and the exchange of energy in radiating systems. Subsequent investigations explored the formation of radiation and the topology of electromagnetic flow in the vicinity of a source~\cite{mikki2013topologicalflow,rfdMikkiMorphogenesis2013}. A scalar formulation of the spectral near-field problem was developed in~\cite{Mikki_and_Anatar_Scalar_NF}, while the study of energy exchange between coupled electromagnetic systems in~\cite{mikki_energy} further illustrated the importance of source geometry and relative position in determining the propagating and non-propagating contributions. These developments suggest a broader use of the spectral decomposition. In the present paper, we extend this line of investigation toward a fundamental theory in which the relative strengths of the partial fields and their dependence on the direction of decomposition provide the basis for a local law of motion.

To construct such a law, we consider a scalar radiation field that is monochromatic in time and generated by a general prescribed spatial source distribution $J(\mathbf x)$, subject to the regularity conditions discussed in Appendix~\ref{sec:rfd_analytical_structure}. The direction along which the field is decomposed at a spatial point $\mathbf x$ is specified by a \emph{local spectral axis} $\hat\alpha=\gamma\hat{x}_3$, where $\gamma\in\mathrm{SO}(3)$ defines the local spectral frame relative to a frame attached to the source description. The position and the spectral frame thus enter the construction together, giving rise to the extended configuration space $\mathcal M=\mathbb R^3\times\mathrm{SO}(3)$ (Section~\ref{sec:rotated_weyl_field_decomposition}). On this space we introduce a dimensionless \emph{propagation potential} $\mathcal V$, defined as the negative ratio of the squared amplitudes of the evanescent and propagating partial fields $\varphi_{\mathrm{ev}}$ and $\varphi_{\mathrm{pr}}$, wherever $\varphi_{\mathrm{pr}}\ne0$ (Definition~\ref{def:rfd_local_propagation_potential}). As this potential approaches its maximum value of zero, the total response approaches a purely propagating partial field in a relative amplitude sense; the associated suppression of coherent interference is quantified in Proposition~\ref{prop:rfd_relative_interference_control}. We use a Riemannian metric $g$ to associate a tangent direction with the variation of $\mathcal V$ on $\mathcal M$ and postulate a unified dynamical law incorporating relaxation and inertia (Eq.~\eqref{eq:rfd_unified_geometric_motion}). On the regular domain of $\mathcal V$, both branches have locally unique solutions under the hypotheses of Proposition~\ref{prop:rfd_local_wellposedness}. The regularity analysis extends the range of systems admitting these dynamics from smooth to H\"older-continuous compactly supported sources and, in exterior observation regions, to general bounded sources of compact support (Propositions~\ref{prop:rfd_smooth_rotated_convolutions} and~\ref{prop:rfd_weak_source_regularity}). These solutions define the \emph{propagaton}, an evolving point-like configuration $q(\tau)=(\mathbf x(\tau),\gamma(\tau))$ carrying a local spectral frame. The massless branch of the law describes ascent of the propagation potential, while the conservative massive branch admits a Lagrangian and Hamiltonian formulation (Section~\ref{sec:rfd_geometric_dynamics}).

The introduction of the spectral frame into the configuration space also brings out a geometrical structure that is important to the dynamics. The partial fields and the propagation potential are covariant under joint Euclidean transformations of the source, observation point and local spectral frame (Proposition~\ref{prop:rfd_joint_source_observer_covariance}). For a fixed source, a continuous symmetry arises from rotations of the transverse basis about the local spectral axis (Proposition~\ref{prop:rfd_axial_rotation_invariance}). With a compatible metric, this symmetry gives rise to an axial momentum that is conserved in the conservative massive branch. The associated geometry makes it possible to reduce the dynamics to a five-dimensional position--axis space. The reduction applies to the massless branch and to inertial motion with vanishing axial momentum (Propositions~\ref{prop:rfd_axial_noether} and~\ref{prop:rfd_five_dimensional_reduction}). The physical implications of the extended configuration space can be seen particularly clearly for an outgoing point source. Although the total radiation field is spherically symmetric, its propagation potential varies with the angle between the local spectral axis and the radial direction, as established analytically in Proposition~\ref{prop:rfd_point_orientation_contrast}. For the metric and initial conditions considered in Section~\ref{sec:rfd_interpretation}, the computed propagaton trajectories depart from the radial direction of conventional field-energy transport. Within the proposed dynamics, the spectral structure of a field can therefore generate a form of motion distinct from the flow of its energy. The metric and dynamical law constitute input data of the theory.

The paper is organized as follows. Section~\ref{sec:rfd_source_response_potential} develops the source-relative spectral decomposition and constructs the propagation potential. Section~\ref{sec:rfd_geometric_dynamics} formulates the unified dynamical law and examines its massless and massive branches. Section~\ref{sec:rfd_mathematical_properties} establishes local well-posedness and develops the symmetry and reduction results. The relational meaning of the theory and the point-source illustration are discussed in Section~\ref{sec:rfd_interpretation}, followed by the conclusions in Section~\ref{sec:rfd_conclusion}. The appendices collect the wave-theoretic identities, analytical regularity results, orientation-dependent examples, quotient geometry and details of the numerical integration. The principal notation is summarized in Table~\ref{tab:rfd_symbols}. We use bold lowercase letters for spatial vectors, nonbold hats for unit vectors, upright Latin capitals for matrices (with $\gamma$ denoting the spectral rotation), $\top$ for transpose, and $\mathrm d$ for differentials and integration measures.

\begin{table}[htbp]
\centering\small
\caption{Principal symbols. $L$, $M$ and $T_0$ denote the formal units of length, of the configuration-space inertia coefficient and of the evolution scale; they are used only to track the dimensions of the dynamical coefficients (Section~\ref{sec:rfd_interpretation}).}
\label{tab:rfd_symbols}
\begin{tabular}{@{}p{0.27\linewidth}p{0.50\linewidth}p{0.17\linewidth}@{}}
\hline
Symbol & Meaning & Dimension\\\hline
$k=\omega/c=2\pi/\lambda$ & wavenumber and wavelength of the monochromatic response & $L^{-1}$, $L$\\
$J,\ \varphi$ & prescribed source phasor and outgoing response phasor & --\\
$G_k^+,\ G_{\mathrm{pr}},\ G_{\mathrm{ev}}$ & outgoing Green function and its propagating and evanescent parts & $L^{-1}$\\
$\gamma,\ \hat\alpha=\gamma\hat{x}_3$ & local spectral frame in $\mathrm{SO}(3)$ and local spectral axis & dimensionless\\
$\mathcal M,\ q=(\mathbf x,\gamma)$ & extended configuration space and one configuration & --\\
$\varphi_{\mathrm{pr}},\ \varphi_{\mathrm{ev}}$ & propagating and evanescent partial fields; $\varphi_{\mathrm{pr}}+\varphi_{\mathrm{ev}}=\varphi$ & as $\varphi$\\
$p,\ e,\ \chi$ & $|\varphi_{\mathrm{pr}}|^2$, $|\varphi_{\mathrm{ev}}|^2$ and their interference $2\operatorname{Re}(\varphi_{\mathrm{pr}}^*\varphi_{\mathrm{ev}})$ & as $|\varphi|^2$\\
$\eta=e/p,\ \mathcal V=-\eta$ & relative evanescent strength and propagation potential & dimensionless\\
$g,\ \ell$ & Riemannian metric on $\mathcal M$; length weighting frame rotations, $\ell=k^{-1}$ & --, $L$\\
$\tau$ & evolution parameter & $T_0$\\
$m_*,\ \nu,\ \Lambda$ & inertia, relaxation and coupling coefficients of the unified dynamical law & $M$, $M/T_0$, $ML^2/T_0^2$\\
$\zeta=\Lambda/\nu$ & mobility of the massless branch & $L^2/T_0$\\
$\mu,\ \vartheta_{\mathrm F}$ & alignment cosine with the spectral axis; angle to the radial flux & dimensionless; degrees\\
$\mathcal J_{\mathrm{ax}}$ & axial momentum & $ML^2/T_0$\\\hline
\end{tabular}
\end{table}

\section{Source--Response Fields and the Propagation Potential}
\label{sec:rfd_source_response_potential}

To develop a mathematical formulation of the source--response relation introduced in Section~1, we consider the generation of a real scalar field $\varphi(\mathbf x,t)$ by a prescribed source field $J(\mathbf x,t)$ in flat spacetime. The scalar formulation allows us to investigate the spectral structure of radiation independently of polarization. We take as our starting point the sourced massless Klein--Gordon equation
\begin{equation}
\label{eq:rfd_sourced_wave_equation}
\nabla^2\varphi(\mathbf x,t)-c^{-2}\partial_t^2\varphi(\mathbf x,t)=-J(\mathbf x,t),
\end{equation}
where $c>0$ is the characteristic wave speed. The variational formulation of Eq.~\eqref{eq:rfd_sourced_wave_equation}, together with its retarded solution and associated energy balance, is given in Appendix~\ref{app:pure_scalar}. For monochromatic fields with time dependence $\exp(-\im\omega t)$, we denote the complex spatial amplitudes of the source and response fields by $J(\mathbf x)$ and $\varphi(\mathbf x)$, respectively, with the physical fields obtained by taking the real parts of the corresponding time-dependent expressions. Equation~\eqref{eq:rfd_sourced_wave_equation} then reduces to the inhomogeneous Helmholtz equation
\begin{equation}
\label{eq:rfd_helmholtz_equation}
(\nabla^2+k^2)\varphi(\mathbf x)=-J(\mathbf x),\qquad k\coloneqq\omega/c>0.
\end{equation}

For a prescribed source distribution $J\in C_c^\infty(\mathbb R^3)$, the space of smooth functions with compact support, let $D_J\coloneqq\operatorname{supp}J$. The Sommerfeld radiation condition (Appendix~\ref{app:pure_scalar}) determines the unique outgoing solution of Eq.~\eqref{eq:rfd_helmholtz_equation} in the form
\begin{equation}
\label{eq:rfd_total_outgoing_response}
\varphi(\mathbf x)=\int_{D_J}\mathrm d^3y\,G_k^+(\mathbf x-\mathbf y)J(\mathbf y),\qquad
G_k^+(\mathbf s)\coloneqq\frac{\exp(\im k|\mathbf s|)}{4\pi|\mathbf s|},
\end{equation}
where the outgoing Green function satisfies $(\nabla^2+k^2)G_k^+=-\delta^{(3)}$ in the distributional sense. Appendix~\ref{sec:rfd_analytical_structure} establishes field and potential regularity for less restrictive source classes, extending the applicability of the subsequent dynamical laws (Proposition~\ref{prop:rfd_weak_source_regularity}). Equation~\eqref{eq:rfd_total_outgoing_response} establishes the spatial relation between the source field $J$ and the radiation field $\varphi$. Our interest in what follows lies in resolving this relation into its propagating and non-propagating contributions and examining their dependence on the local direction of spectral decomposition.

\subsection{Local spectral frames and partial fields}
\label{sec:rotated_weyl_field_decomposition}

The spectral representation of the outgoing Green function $G_k^+$ separates the radiation field into propagating and evanescent contributions relative to a selected spatial direction. For a transverse wavevector $\mathbf k_{\mathrm t}$, the longitudinal wavenumber $k_{\mathrm l}=\sqrt{k^2-|\mathbf k_{\mathrm t}|^2+\im0}$ determines the character of the corresponding mode. Propagating modes occupy the spectral disk $|\mathbf k_{\mathrm t}|^2<k^2$, whereas evanescent modes occupy its exterior, $|\mathbf k_{\mathrm t}|^2>k^2$. The corresponding Green kernels $G_{\mathrm{pr}}$ and $G_{\mathrm{ev}}$, whose explicit spectral representations follow from the Weyl expansion (Eqs.~\eqref{eq:G_pr} and~\eqref{eq:G_ev} in Appendix~\ref{app:weyl}), satisfy $G_k^+=G_{\mathrm{pr}}+G_{\mathrm{ev}}$; the grazing circle is understood by the limiting prescription in Eq.~\eqref{eq:k3_branch}.

Let $\{\hat{x}_1,\hat{x}_2,\hat{x}_3\}$ be a global Cartesian basis attached to the prescribed source description. We introduce a \emph{local spectral frame} through a rotation $\gamma\in\mathrm{SO}(3)$, with $\gamma^\top\gamma=\mathrm I_3$ and $\det\gamma=1$, defining the rotated basis $\hat{x}'_j=\gamma\hat{x}_j$. The unit vector $\hat\alpha=\gamma\hat{x}_3$, referred to henceforth as the \emph{local spectral axis}, specifies the longitudinal direction relative to which the Weyl decomposition is performed. For a source point $\mathbf y$ and an observation point $\mathbf x$, the displacement $\mathbf s=\mathbf x-\mathbf y$ admits the local coordinate representation
\begin{equation}
\label{eq:rfd_rotated_coordinates}
\mathbf s_\gamma=\gamma^{\top}(\mathbf x-\mathbf y),\qquad
z_\gamma=\hat{x}_3\cdot\mathbf s_\gamma=\hat\alpha\cdot(\mathbf x-\mathbf y),\qquad
\mathbf s_{\mathrm t,\gamma}=\mathbf s_\gamma-z_\gamma\hat{x}_3.
\end{equation}
The quantities $z_\gamma$ and $\mathbf s_{\mathrm t,\gamma}$ determine the longitudinal and transverse separations entering the Weyl representation. Consequently, for $\iota\in\{\mathrm{pr},\mathrm{ev}\}$, the rotated Green kernels $G_\iota(\mathbf s;\gamma)\coloneqq G_\iota(\gamma^\top\mathbf s)$ generate the propagating and evanescent \emph{partial fields}
\begin{equation}
\label{eq:rfd_rotated_source_field_integrals}
\varphi_\iota(\mathbf x,\gamma)=\int_{D_J}\mathrm d^3y\,J(\mathbf y)G_\iota(\mathbf x-\mathbf y;\gamma).
\end{equation}
The rotational invariance of $G_k^+$, together with its spectral decomposition, gives the exact reconstruction
\begin{equation}
\label{eq:rfd_total_partial_reconstruction}
\varphi_{\mathrm{pr}}(\mathbf x,\gamma)+\varphi_{\mathrm{ev}}(\mathbf x,\gamma)=\varphi(\mathbf x).
\end{equation}
Thus a change in the local spectral frame generally redistributes the contributions of the two partial fields to the total radiation field $\varphi$. The explicit integral representations of the rotated kernels follow from the azimuthal integration of the Weyl expansion (Appendix~\ref{app:weyl}):
\begin{align}
G_{\mathrm{pr}}(\mathbf x-\mathbf y;\gamma)
&=\frac{\im}{4\pi}\int_0^k\mathrm d\sigma\,
J_0\!\left(|\mathbf s_{\mathrm t,\gamma}|\sqrt{k^2-\sigma^2}\right)
 e^{\im\sigma|z_\gamma|},
\label{eq:rfd_rotated_propagating_kernel}\\
G_{\mathrm{ev}}(\mathbf x-\mathbf y;\gamma)
&=\frac{1}{4\pi}\int_0^\infty\mathrm d\sigma\,
J_0\!\left(|\mathbf s_{\mathrm t,\gamma}|\sqrt{k^2+\sigma^2}\right)
 e^{-\sigma|z_\gamma|},
\label{eq:rfd_rotated_evanescent_kernel}
\end{align}
where $J_0$ is the Bessel function of the first kind of order zero and $\sigma$ is a longitudinal spectral parameter.\footnote{At $z_\gamma=0$ with $|\mathbf s_{\mathrm t,\gamma}|>0$, the evanescent integral denotes its Abel or distributional boundary value, as in Appendix~\ref{app:weyl}. The coincident-source singularity occurs at $\mathbf x=\mathbf y$.} For spatially extended sources, Eq.~\eqref{eq:rfd_rotated_source_field_integrals} incorporates contributions from source points on both sides of the transverse plane passing through $\mathbf x$. The ordinary one-sided angular-spectrum representation is recovered outside the longitudinal interval $\min_{\mathbf y\in D_J}\hat\alpha\cdot\mathbf y\leq\hat\alpha\cdot\mathbf x\leq\max_{\mathbf y\in D_J}\hat\alpha\cdot\mathbf y$.

Equations~\eqref{eq:rfd_rotated_propagating_kernel} and~\eqref{eq:rfd_rotated_evanescent_kernel} reveal two geometrical properties of the spectral partition. First, the dependence of the kernels on the transverse displacement through its magnitude $|\mathbf s_{\mathrm t,\gamma}|$ gives rise to a residual axial symmetry: rotations of the transverse basis about $\hat\alpha$ preserve the partial fields for arbitrary admissible source distributions (Proposition~\ref{prop:rfd_axial_rotation_invariance}). Second, the partial fields generally vary with the direction of $\hat\alpha$, as demonstrated analytically and numerically in Appendix~\ref{app:rfd_point_source_rotation}. To understand the relational significance of this directional dependence, consider a simultaneous rotation and translation of the source field, the observation point, and the local spectral frame. The following proposition establishes the covariance of the spectral partition under such transformations.

\begin{proposition}[Joint Euclidean covariance of the source--response partition]
\label{prop:rfd_joint_source_observer_covariance}
Let $\mathrm R\in\mathrm{SO}(3)$ and $\mathbf b\in\mathbb R^3$, and define the transformed source field by $J_{\mathrm R,\mathbf b}(\mathbf y)=J(\mathrm R^\top(\mathbf y-\mathbf b))$. Then, for each $\iota\in\{\mathrm{pr},\mathrm{ev}\}$,
\begin{equation}
\label{eq:rfd_joint_covariance}
\varphi_{\iota,J_{\mathrm R,\mathbf b}}(\mathrm R\mathbf x+\mathbf b,\mathrm R\gamma)
=\varphi_{\iota,J}(\mathbf x,\gamma),\qquad
\mathcal V_{J_{\mathrm R,\mathbf b}}(\mathrm R\mathbf x+\mathbf b,\mathrm R\gamma)
=\mathcal V_J(\mathbf x,\gamma),
\end{equation}
where $\mathcal V_J$ is the propagation potential introduced in Definition~\ref{def:rfd_local_propagation_potential}, and the second identity holds on the corresponding regular domains.
\end{proposition}
\begin{proof}
Performing the change of variables $\mathbf y'=\mathrm R\mathbf y+\mathbf b$ in Eq.~\eqref{eq:rfd_rotated_source_field_integrals} preserves the volume element and gives $(\mathrm R\gamma)^\top[(\mathrm R\mathbf x+\mathbf b)-(\mathrm R\mathbf y+\mathbf b)]=\gamma^\top(\mathbf x-\mathbf y)$. The invariance of the partial fields follows directly, together with that of their squared-amplitude quotient wherever its propagating denominator is nonzero.
\end{proof}

The significance of Proposition~\ref{prop:rfd_joint_source_observer_covariance} lies in establishing that the spectral structure of the radiation field is determined by the relative geometrical arrangement of the source $J$, the point $\mathbf x$, and the local spectral frame $\gamma$. The covariance extends to the dynamical laws under the metric compatibility conditions established in Section~\ref{sec:rfd_symmetry_structure}. Furthermore, the dependence of the kernels on $|\mathbf s|$ and $|\hat\alpha\cdot\mathbf s|$ extends the covariance to orthogonal transformations $\mathrm R\in\mathrm O(3)$ acting as $(\mathbf x,\hat\alpha)\mapsto(\mathrm R\mathbf x+\mathbf b,\mathrm R\hat\alpha)$ on the position--axis space $\mathbb R^3\times S^2$; reflections are used in Appendix~\ref{app:rfd_numerical_dynamics}.

\begin{figure}[!tbp]
\centering
\includegraphics[width=0.92\linewidth]{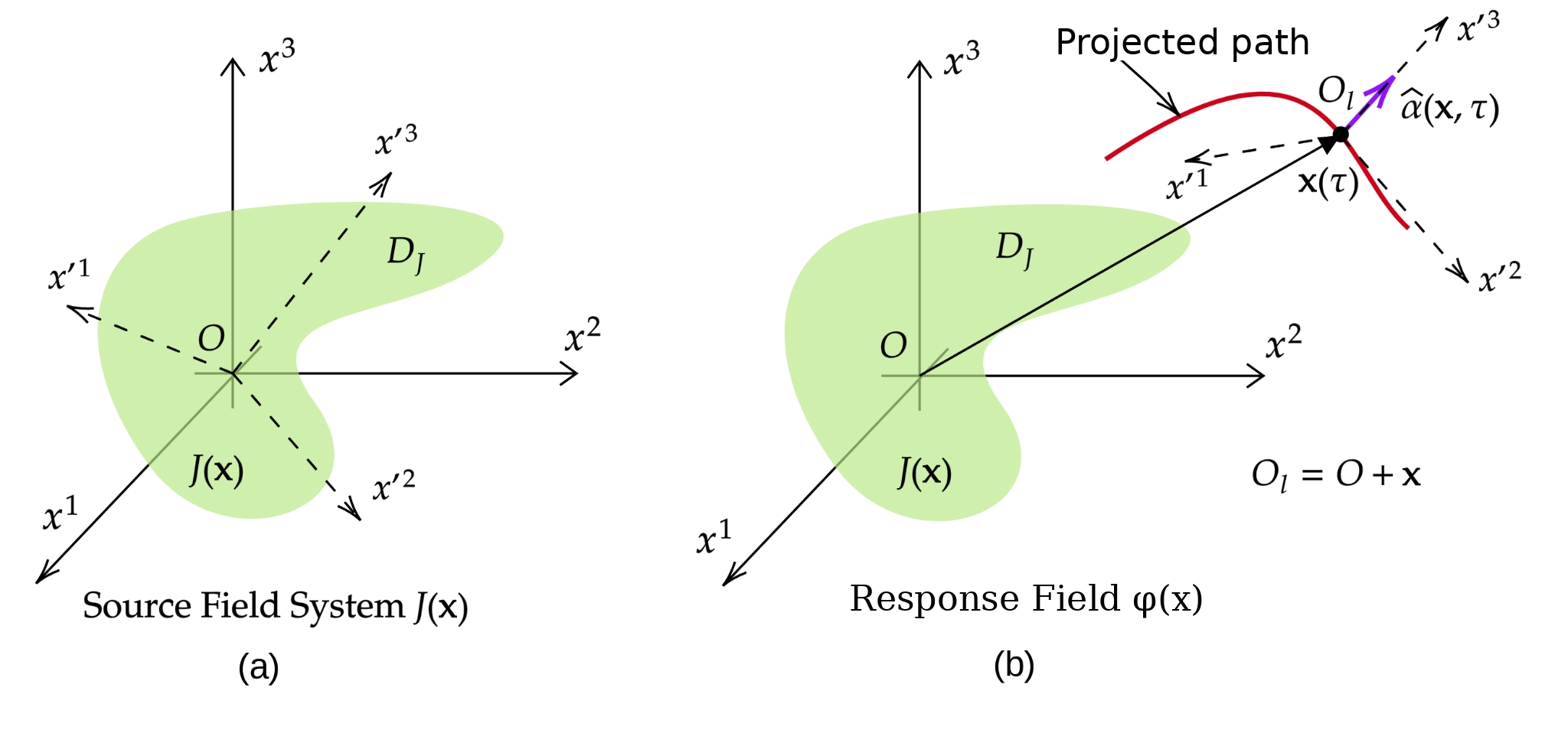}
\caption{Source--response geometry and the local spectral frame. (a) Global Cartesian basis attached to the source field $J$, with support $D_J$, and the rotated spectral basis defined by $\gamma$. (b) Evolution of the local spectral frame along the schematic, spatially projected propagaton trajectory $\mathbf x(\tau)$. The local spectral axis is given by $\hat\alpha=\gamma(\tau)\hat{x}_3$.}
\label{fig:rfd_source_anchored_spectral_frame}
\end{figure}

These considerations suggest treating the spatial point $\mathbf x$ and its local spectral frame $\gamma$ as independent configuration variables and introducing the six-dimensional \emph{extended configuration space}
\begin{equation}
\label{eq:rfd_extended_configuration_space}
\mathcal M\coloneqq\mathbb R^3\times\mathrm{SO}(3),\qquad q=(\mathbf x,\gamma)\in\mathcal M.
\end{equation}
The geometry of $\mathcal M$ combines ordinary position space (a fixed-frame submanifold) with the non-Euclidean rotational degrees of freedom of the local spectral frame. As illustrated in Fig.~\ref{fig:rfd_source_anchored_spectral_frame}, a curve $q(\tau)=(\mathbf x(\tau),\gamma(\tau))$ describes the simultaneous evolution of the spatial position and the local frame relative to the prescribed source field. To formulate a law governing this evolution, we introduce in the next section a scalar function on $\mathcal M$ characterizing the relative strengths of the partial fields $\varphi_{\mathrm{pr}}$ and $\varphi_{\mathrm{ev}}$ and controlling their coherent interference.

\subsection{The local propagation potential}
\label{sec:rfd_local_propagation_potential}

The spectral decomposition developed in Section~\ref{sec:rotated_weyl_field_decomposition} provides a means of characterizing the local state of propagation of the radiation field $\varphi$ relative to the source field $J$. At each spatial point $\mathbf x$, the relative strengths of the partial fields $\varphi_{\mathrm{pr}}$ and $\varphi_{\mathrm{ev}}$ depend on the direction of the local spectral axis $\hat\alpha$. To incorporate this dependence into a theory of motion, we seek a scalar function on the extended configuration space $\mathcal M$ whose variation characterizes changes in the relative composition of the radiation field. For a configuration $q=(\mathbf x,\gamma)\in\mathcal M$, introduce the squared amplitudes and their coherent interaction term
\begin{equation}
\label{eq:rfd_partial_amplitudes}
p(q)=|\varphi_{\mathrm{pr}}(q)|^2,\quad
e(q)=|\varphi_{\mathrm{ev}}(q)|^2,\quad
\chi(q)=2\operatorname{Re}[\varphi_{\mathrm{pr}}(q)^*\varphi_{\mathrm{ev}}(q)].
\end{equation}
The exact reconstruction of the radiation field in Eq.~\eqref{eq:rfd_total_partial_reconstruction} gives $|\varphi(\mathbf x)|^2=p+e+\chi$, where $\chi$ represents the signed interference between the propagating and evanescent partial fields, analogous to the interaction contributions encountered in spectral near-field theory~\cite{mikki_nf1,mikki_nf2}. These quantities characterize the coherent amplitude structure of the radiation field; its canonical energy density is discussed separately in Appendix~\ref{app:pure_scalar}. A dimensionless measure of the relative strength of the evanescent contribution is provided by the ratio $\eta=e/p$ where $p>0$. Assigning a negative sign to this ratio yields a scalar whose increase corresponds to a decreasing relative evanescent contribution, leading to the following definition.

\begin{definition}[Local propagation potential]
\label{def:rfd_local_propagation_potential}
On the open regular domain $\operatorname{Dom}(\mathcal V)=\{q\in\mathcal M:p(q)>0\}$, we define the local propagation potential by
\begin{equation}
\label{eq:rfd_propagation_potential_ratio}
\eta(q)=\frac{e(q)}{p(q)},\qquad
\mathcal V(q)=-\eta(q)
=-\frac{|\varphi_{\mathrm{ev}}(q)|^2}{|\varphi_{\mathrm{pr}}(q)|^2}.
\end{equation}
\end{definition}

The potential satisfies $\mathcal V\leq0$, with equality precisely at regular configurations where $\varphi_{\mathrm{ev}}=0$. By virtue of the linearity of the source--response relation, $\mathcal V$ is invariant under a nonzero complex rescaling of the source field $J$ and possesses the joint Euclidean covariance established in Proposition~\ref{prop:rfd_joint_source_observer_covariance}. Its dependence on the local spectral axis is demonstrated analytically for a point source and numerically for an asymmetric source distribution in Appendix~\ref{app:rfd_point_source_rotation}. From Eq.~\eqref{eq:rfd_total_partial_reconstruction}, we have $|\varphi-\varphi_{\mathrm{pr}}|/|\varphi_{\mathrm{pr}}|=\sqrt{-\mathcal V}$, so that the square root of the negative potential measures the relative amplitude discrepancy between the total radiation field and its propagating partial field. Of particular interest for the dynamical construction is the behavior of the coherent interaction $\chi$ as the propagation potential approaches zero. The following proposition establishes quantitative bounds on this interaction and on the relative contribution of the propagating partial field to the total response.

\begin{proposition}[Control of coherent interaction by the propagation potential]
\label{prop:rfd_relative_interference_control}
At every regular configuration,
\begin{equation}
\label{eq:rfd_interference_relative_bound}
\frac{|\chi|}{p}\leq2\sqrt\eta=2\sqrt{-\mathcal V},\qquad
(1-\sqrt\eta)^2\leq\frac{|\varphi|^2}{p}\leq(1+\sqrt\eta)^2.
\end{equation}
For $0\leq\eta<1$, the total squared amplitude is strictly positive and satisfies
\begin{equation}
\label{eq:rfd_coherent_purity_bounds}
\frac{1}{(1+\sqrt\eta)^2}\leq\frac{p}{|\varphi|^2}\leq\frac{1}{(1-\sqrt\eta)^2}.
\end{equation}
Consequently, along regular configurations, $\mathcal V\to0$ implies $\chi/p\to0$ and $|\varphi|^2/p\to1$.
\end{proposition}
\begin{proof}
The first bound follows from the estimate $|\chi|\leq2|\varphi_{\mathrm{pr}}||\varphi_{\mathrm{ev}}|=2\sqrt{pe}$. Applying the triangle and reverse-triangle inequalities to the coherent sum $\varphi=\varphi_{\mathrm{pr}}+\varphi_{\mathrm{ev}}$ and dividing by $p>0$ gives the remaining inequalities in Eq.~\eqref{eq:rfd_interference_relative_bound}. For $\eta<1$, inversion of these bounds yields Eq.~\eqref{eq:rfd_coherent_purity_bounds}. The limiting relations follow by taking $\eta\to0$.
\end{proof}

The physical significance of Proposition~\ref{prop:rfd_relative_interference_control} is that the increase of the propagation potential toward zero is accompanied by a relative suppression of both the evanescent contribution and its coherent interaction with the propagating partial field. In this sense, $\mathcal V$ characterizes the degree of relative propagating purity of the local field decomposition. To investigate how this property varies along curves in the extended configuration space $\mathcal M$, we consider the differential of the potential,
\begin{equation}
\label{eq:rfd_propagation_potential_differential}
\mathrm d\mathcal V=\frac{e}{p^2}\,\mathrm dp-\frac{1}{p}\,\mathrm de.
\end{equation}
For a tangent vector $v\in T_q\mathcal M$, the condition $\mathrm d\mathcal V_q[v]>0$ is equivalent to a first-order decrease of the relative evanescent strength $\eta$ along $v$. The analytical properties of this differential, including its regularity for distributed sources and the behavior of $\mathcal V$ at zeros of the partial fields and the total response, are investigated in Appendix~\ref{sec:rfd_analytical_structure}. The construction of a local law of motion from $\mathrm d\mathcal V$ requires a geometric structure relating covectors to tangent vectors on $\mathcal M$. This leads to the Riemannian formulation of propagaton dynamics developed in Section~\ref{sec:rfd_geometric_dynamics}.

\section{Geometric Dynamics and the Emergence of Propagatons}
\label{sec:rfd_geometric_dynamics}

The propagation potential $\mathcal V$ introduced in Section~\ref{sec:rfd_local_propagation_potential} characterizes the relative propagating content of the radiation field at each configuration $q=(\mathbf x,\gamma)$. To construct a theory of motion from this scalar, we must first specify the geometry of the extended configuration space $\mathcal M=\mathbb R^3\times\mathrm{SO}(3)$. The translational and rotational degrees of freedom of $\mathcal M$ possess different geometrical structures: whereas $\mathbb R^3$ carries its natural Euclidean metric, the rotation group $\mathrm{SO}(3)$ is a compact, non-Euclidean manifold admitting a variety of Riemannian metrics. Moreover, the geometry of the product space may incorporate coupling between spatial displacements and changes in the local spectral frame. We therefore equip the regular domain of $\mathcal V$ in $\mathcal M$ with a positive-definite Riemannian metric $g$, which determines the inner product of tangent vectors at each configuration. The differential $\mathrm d\mathcal V_q\in T_q^*\mathcal M$ is then associated with a tangent vector $\operatorname{grad}_g\mathcal V(q)\in T_q\mathcal M$ through the metric duality relations
\begin{equation}
\label{eq:rfd_riemannian_gradient_definition}
g_q(\operatorname{grad}_g\mathcal V,v)=\mathrm d\mathcal V_q[v],\qquad
(\operatorname{grad}_g\mathcal V)^a=g^{ab}\partial_b\mathcal V,
\end{equation}
where $v\in T_q\mathcal M$, $q^a$ are local coordinates on $\mathcal M$, and $g^{ab}$ are the components of the inverse metric. The choice of $g$ thus determines how variations in the relative strengths of the propagating and evanescent partial fields are translated into directions of motion in the extended configuration space. The geometrical properties of $g$ also enter the inertial dynamics developed in Section~\ref{sec:rfd_massive_inertial_branch} through its associated Levi--Civita connection.

A particularly useful choice, which we employ in the numerical investigations of Section~\ref{sec:rfd_interpretation}, is the source-independent product metric $g_0=\mathrm d\mathbf x\cdot\mathrm d\mathbf x+\ell^2\langle\gamma^{-1}\mathrm d\gamma,\gamma^{-1}\mathrm d\gamma\rangle$, where $\gamma^{-1}\mathrm d\gamma$ takes values in the Lie algebra $\mathfrak{so}(3)$, equipped with the adjoint-invariant inner product $\langle\mathrm A,\mathrm B\rangle=\tfrac12\operatorname{tr}(\mathrm A^{\top}\mathrm B)$, which induces a bi-invariant metric on $\mathrm{SO}(3)$. The characteristic length $\ell>0$ assigns a relative geometrical weight to changes in the local spectral frame and ensures dimensional consistency between the translational and rotational contributions to the metric. To describe the rotational geometry explicitly, we associate with each vector $\mathbf a=(a_1,a_2,a_3)^{\top}\in\mathbb R^3$ the antisymmetric matrix $\mathrm A(\mathbf a)\in\mathfrak{so}(3)$, defined by $\mathrm A(\mathbf a)\mathbf b=\mathbf a\times\mathbf b$ for every $\mathbf b\in\mathbb R^3$, i.e.,
\[
\mathrm A(\mathbf a)=\begin{pmatrix}0&-a_3&a_2\\a_3&0&-a_1\\-a_2&a_1&0\end{pmatrix}.
\]
The induced Lie-algebra norm satisfies $\|\mathrm A(\mathbf a)\|_{\mathfrak{so}(3)}^2=\tfrac12\operatorname{tr}[\mathrm A(\mathbf a)^{\top}\mathrm A(\mathbf a)]=|\mathbf a|^2$. This normalization also makes the natural projection $\gamma\mapsto\gamma\hat{x}_3$ from $\mathrm{SO}(3)$ onto the unit sphere $S^2$ a Riemannian submersion for the unscaled rotational metric, a property used in the geometrical reduction developed in Section~\ref{sec:rfd_axial_momentum_reduction} and Appendix~\ref{app:rfd_quotient_technical}. The product metric $g_0$ is invariant under joint Euclidean transformations and residual axial frame rotations, thereby satisfying the metric compatibility conditions for dynamical covariance established in Section~\ref{sec:rfd_symmetry_structure}. In the following development, we formulate the dynamical laws for a general metric $g$, allowing the geometry of the extended configuration space to constitute an independent element of the theory.

\subsection{A unified dynamical law and its massless branch}
\label{sec:rfd_unified_dynamics}

The propagation potential $\mathcal V$ and the Riemannian metric $g$ provide the basic ingredients for constructing a local theory of motion on the extended configuration space $\mathcal M$. At regular configurations with nonzero differential, the metric gradient $\operatorname{grad}_g\mathcal V$ determines the direction of steepest increase of the propagation potential, suggesting its use as the driving term in a dynamical law governing the evolution of both the spatial position and the local spectral frame. We introduce the \emph{propagaton} as a point-like configuration $q(\tau)=(\mathbf x(\tau),\gamma(\tau))$ whose evolution is governed by the spectral structure of the radiation field relative to its source. To accommodate both inertial and relaxational mechanisms of motion within a single formulation, we postulate the following unified dynamical law:
\begin{equation}
\label{eq:rfd_unified_geometric_motion}
m_*\nabla^g_{\dot q}\dot q+\nu\dot q
=\Lambda\operatorname{grad}_g\mathcal V(q),
\end{equation}
where $m_*,\nu\geq0$, $(m_*,\nu)\ne(0,0)$, $\Lambda>0$ determines the strength of the potential driving term, and $\nabla^g$ is the Levi--Civita connection associated with $g$. The coefficients $m_*$ and $\nu$ characterize the inertial and relaxational properties of the propagaton, respectively. The evolution parameter $\tau$ parametrizes the motion on $\mathcal M$. The term $\nu\dot q$ describes relaxation within the extended configuration space, whose energetic significance follows from the configuration-space energy balance (Eq.~\eqref{eq:rfd_configuration_energy_identity}) derived in Section~\ref{sec:rfd_massive_inertial_branch}. The radiation field $\varphi$ itself obeys a separate energy balance, given in Appendix~\ref{app:pure_scalar}.

For $m_*=0$ and $\nu>0$, Eq.~\eqref{eq:rfd_unified_geometric_motion} reduces exactly to the \emph{massless gradient branch} of propagaton dynamics,
\begin{equation}
\label{eq:rfd_massless_propagaton_motion}
\dot q=\zeta\operatorname{grad}_g\mathcal V(q),
\qquad \zeta\coloneqq\Lambda/\nu>0,
\end{equation}
where $\zeta$ is the mobility coefficient. An immediate consequence of Eq.~\eqref{eq:rfd_massless_propagaton_motion}, obtained using the metric duality relation~\eqref{eq:rfd_riemannian_gradient_definition} and the definition $\eta=-\mathcal V$, is
\begin{equation}
\label{eq:rfd_massless_lyapunov_identity}
\frac{\mathrm d\mathcal V}{\mathrm d\tau}
=\zeta\|\operatorname{grad}_g\mathcal V\|_g^2\geq0,
\qquad \frac{\mathrm d\eta}{\mathrm d\tau}
=-\zeta\|\operatorname{grad}_g\mathcal V\|_g^2\leq0.
\end{equation}
Thus the relative propagating purity is nondecreasing along a regular massless propagaton trajectory, with strict increase of $\mathcal V$ away from its critical configurations. In view of Proposition~\ref{prop:rfd_relative_interference_control}, whenever $\mathcal V$ approaches zero along a trajectory, the evolution is accompanied by a relative suppression of both the evanescent partial field $\varphi_{\mathrm{ev}}$ and its coherent interaction with the propagating partial field $\varphi_{\mathrm{pr}}$. This property establishes a direct relationship between propagaton motion and the relative spectral composition of the radiation field. The local existence and continuation properties of the massless dynamical law are investigated in Section~\ref{sec:rfd_dynamics_regularity}, while its behavior for a point-source radiation field is illustrated numerically in Section~\ref{sec:rfd_interpretation}.

The massless dynamical law also admits a variational characterization in terms of the instantaneous tangent velocity. At a fixed regular configuration $q$, consider the quadratic functional $\mathscr R_q:T_q\mathcal M\to\mathbb R$, comprising a metric-dependent quadratic cost associated with the tangent velocity and a linear contribution associated with the variation of the propagation potential. Its definition and unique minimizer are given by
\begin{equation}
\label{eq:rfd_massless_velocity_variation}
\mathscr R_q(v)\coloneqq\frac{\nu}{2}g_q(v,v)-\Lambda\,\mathrm d\mathcal V_q[v],
\qquad
\operatorname*{arg\,min}_{v\in T_q\mathcal M}\mathscr R_q(v)
=\zeta\operatorname{grad}_g\mathcal V(q).
\end{equation}
Indeed, stationarity with respect to $v$ gives $\nu g_q(v,w)=\Lambda\,\mathrm d\mathcal V_q[w]$ for every $w\in T_q\mathcal M$, which, by Eq.~\eqref{eq:rfd_riemannian_gradient_definition}, yields the massless velocity~\eqref{eq:rfd_massless_propagaton_motion}. The positive definiteness of $g$ and the condition $\nu>0$ ensure that the stationary point is the unique global minimizer. Equation~\eqref{eq:rfd_massless_velocity_variation} therefore provides an equivalent formulation of the massless branch through an instantaneous variational principle on the tangent space.

For either dynamical branch, the spatial trajectory of the propagaton is obtained by projecting $q(\tau)$ onto $\mathbb R^3$, giving the curve $\mathbf x(\tau)$ with projected velocity $\mathbf v_\tau=\mathrm d\mathbf x/\mathrm d\tau$. The relationship between the direction of spatial motion and the local spectral axis $\hat\alpha=\gamma\hat{x}_3$ can be characterized by the dimensionless alignment cosine $\mu(\tau)=\hat\alpha(\tau)\cdot\hat v_\tau(\tau)\in[-1,1]$, where $\hat v_\tau=\mathbf v_\tau/|\mathbf v_\tau|$ is defined for nonzero projected velocities. This quantity, together with the angle between the projected velocity and the conventional field-energy flux, is investigated numerically in Section~\ref{sec:rfd_interpretation}.

\subsection{The massive inertial branch}
\label{sec:rfd_massive_inertial_branch}

The massless branch developed in Section~\ref{sec:rfd_unified_dynamics} describes an instantaneous response of the propagaton to the metric gradient of the propagation potential. The introduction of inertia extends this description by allowing the evolution to depend on both the configuration $q$ and its tangent velocity $\dot q$. For $m_*>0$ and $\nu=0$, the unified dynamical law~\eqref{eq:rfd_unified_geometric_motion} takes the form
\begin{equation}
\label{eq:rfd_massive_geometric_equation}
m_*\nabla^g_{\dot q}\dot q=\Lambda\operatorname{grad}_g\mathcal V(q).
\end{equation}
This equation defines the \emph{massive inertial branch} of propagaton dynamics. Its geometrical structure is that of a natural mechanical system on a Riemannian configuration space~\cite{arnold1989mathematical_methods_of_classical_mechanics,MarsdenRatiu1999}, with kinetic energy determined by the metric $g$ and potential energy given by $-\Lambda\mathcal V$. The corresponding Lagrangian and action functional are
\begin{equation}
\label{eq:rfd_massive_lagrangian}
L_{m_*}(q,\dot q)=\frac{m_*}{2}g_q(\dot q,\dot q)+\Lambda\mathcal V(q),\qquad
\mathcal S_{m_*}[q]=\int_{\tau_0}^{\tau_1}\mathrm d\tau\,L_{m_*}(q,\dot q).
\end{equation}
Stationarity of the action $\mathcal S_{m_*}$ under fixed-endpoint variations yields the covariant equation of motion~\eqref{eq:rfd_massive_geometric_equation}. In local coordinates $q^a$, the Euler--Lagrange equation reads $m_*(\ddot q^a+\Gamma^a_{bc}\dot q^b\dot q^c)=\Lambda g^{ab}\partial_b\mathcal V$, where $\Gamma^a_{bc}$ are the Christoffel symbols of the Levi--Civita connection associated with $g$. The coefficient $m_*$ therefore plays the role of an inertia associated with motion in the extended configuration space, encompassing both spatial displacement and rotation of the local spectral frame. This inertia is a dynamical parameter of the propagaton, while the underlying scalar field in Eq.~\eqref{eq:rfd_sourced_wave_equation} is massless.

The Lagrangian formulation also provides a natural Hamiltonian description of the massive propagaton. Introducing the canonical momentum $\pi_a=\partial L_{m_*}/\partial\dot q^a=m_*g_{ab}\dot q^b$, the positive definiteness of $g$ and the condition $m_*>0$ ensure that the Legendre transformation from $T\operatorname{Dom}(\mathcal V)$ to $T^*\operatorname{Dom}(\mathcal V)$ is invertible. The corresponding Hamiltonian on the cotangent bundle is
\begin{equation}
\label{eq:rfd_massive_hamiltonian}
\mathscr H(q,\pi)=\frac{1}{2m_*}g^{ab}\pi_a\pi_b-\Lambda\mathcal V(q).
\end{equation}
The massive propagaton thus admits the standard symplectic formulation of geometric mechanics, with the propagation potential supplying the interaction with the source-generated radiation field. The symmetry properties of this interaction give rise to a conserved axial momentum under appropriate metric conditions, as established in Proposition~\ref{prop:rfd_axial_noether}, and provide the basis for the reduction of inertial dynamics with vanishing axial momentum to a lower-dimensional configuration space (Proposition~\ref{prop:rfd_five_dimensional_reduction}).

The relationship between the massive and massless branches becomes particularly transparent through an energy balance associated with the unified dynamical law~\eqref{eq:rfd_unified_geometric_motion}. For an arbitrary regular solution of this law, introduce the configuration-space energy $\mathcal E_{\mathrm{cfg}}=\tfrac{m_*}{2}\|\dot q\|_g^2-\Lambda\mathcal V$. Taking its derivative along the trajectory and using metric compatibility of the Levi--Civita connection gives
\begin{equation}
\label{eq:rfd_configuration_energy_identity}
\frac{\mathrm d\mathcal E_{\mathrm{cfg}}}{\mathrm d\tau}=-\nu\|\dot q\|_g^2\leq0.
\end{equation}
For the conservative massive branch, $\mathcal E_{\mathrm{cfg}}$ coincides with the Hamiltonian~\eqref{eq:rfd_massive_hamiltonian} evaluated at the canonical momentum and is conserved along the trajectory. Changes in the propagation potential are then accompanied by an exchange between kinetic and potential energy. For the massless branch, $\mathcal E_{\mathrm{cfg}}=-\Lambda\mathcal V$, and Eq.~\eqref{eq:rfd_configuration_energy_identity} recovers the monotonicity relations~\eqref{eq:rfd_massless_lyapunov_identity}. When both inertia and relaxation are present, $\mathcal E_{\mathrm{cfg}}$ is nonincreasing, while the propagation potential can exhibit intervals of ascent and descent. The nonnegativity of the kinetic energy and of $-\mathcal V=\eta$ also makes Eq.~\eqref{eq:rfd_configuration_energy_identity} useful in establishing restrictions on the configurations accessible to a propagaton (Proposition~\ref{prop:rfd_regular_domain_barrier}). The numerical manifestations of these energetic properties are examined in Section~\ref{sec:rfd_interpretation} and Appendix~\ref{app:rfd_numerical_dynamics}; the energy balance of the scalar field itself is given in Appendix~\ref{app:pure_scalar}.

The relative strengths of the inertial, relaxational, and potential driving contributions to Eq.~\eqref{eq:rfd_unified_geometric_motion} are characterized by the ratios $m_*/\Lambda$ and $\nu/\Lambda$. Introducing a reference length $L_0$ and an evolution scale $T_0$, we obtain the dimensionless coefficients $\bar m=m_*L_0^2/(\Lambda T_0^2)$ and $\bar\nu=\nu L_0^2/(\Lambda T_0)$. For $\nu>0$, the choice $T_0=\nu L_0^2/\Lambda$ gives $\bar\nu=1$ and $\bar m=m_*\Lambda/(\nu^2L_0^2)$, thereby expressing the relative importance of inertia through a single dimensionless coefficient at fixed normalized geometry. The general dimensionless coefficients are used with $L_0=\lambda$ in the numerical formulation of Section~\ref{sec:rfd_interpretation} and Appendix~\ref{app:rfd_numerical_dynamics}. The geometry itself possesses a related scaling freedom: a constant rescaling $g\mapsto c_g g$, with $c_g>0$, preserves the Levi--Civita connection and transforms the metric gradient according to $\operatorname{grad}_{c_g g}\mathcal V=c_g^{-1}\operatorname{grad}_g\mathcal V$. The simultaneous transformation $\Lambda\mapsto c_g\Lambda$ therefore preserves the dynamical law. In the present formulation, the metric $g$ determines both the inertial geometry and the isotropic relaxation represented by $\nu\dot q$.\footnote{A separate positive-definite relaxation metric $\mathsf r$ would give the relaxation tensor $g^{-1}\mathsf r$ and a first-order gradient with respect to $\mathsf r$.}

\section{Analytical and Geometrical Properties of Propagaton Dynamics}
\label{sec:rfd_mathematical_properties}

The dynamical formulation developed in Section~\ref{sec:rfd_geometric_dynamics} raises two mathematical questions concerning the existence of propagaton trajectories and the influence of spectral symmetries on their evolution. The first is intimately related to the regularity of the propagation potential $\mathcal V$, which in turn depends on the analytical properties of the partial fields $\varphi_{\mathrm{pr}}$ and $\varphi_{\mathrm{ev}}$. Convolution of the Weyl kernels with smooth, compactly supported sources $J$ yields smooth partial fields on $\mathcal M$ and hence a smooth propagation potential on its regular domain $\operatorname{Dom}(\mathcal V)$ (Proposition~\ref{prop:rfd_smooth_rotated_convolutions}). The extension of these regularity results to less restrictive source classes is important for the physical applicability of the theory: it enlarges the range of systems for which the propagaton dynamical laws admit locally unique solutions (Proposition~\ref{prop:rfd_weak_source_regularity}). These analytical results provide the basis for investigating the existence, uniqueness, and continuation of propagaton trajectories in Section~\ref{sec:rfd_dynamics_regularity}. The rotational structure of the spectral partition gives rise to a further set of geometrical properties. In Section~\ref{sec:rfd_symmetry_structure}, we establish how the intrinsic axial symmetry of the partial fields, together with appropriate symmetries of the metric $g$, governs the equivariance of the dynamical laws. Section~\ref{sec:rfd_axial_momentum_reduction} develops the resulting axial momentum law and geometric reduction of the extended configuration space.

\subsection{Local existence and continuation}
\label{sec:rfd_dynamics_regularity}

The local solvability of the unified dynamical law~\eqref{eq:rfd_unified_geometric_motion} is determined by the regularity of the propagation potential $\mathcal V$ and the configuration-space metric $g$. The massless branch defines a vector field on the regular domain $\operatorname{Dom}(\mathcal V)$, whereas the inertial dynamics defines a vector field on its tangent bundle. The analytical results established in Propositions~\ref{prop:rfd_smooth_rotated_convolutions} and~\ref{prop:rfd_weak_source_regularity} provide sufficient regularity for both formulations, leading to the following existence and uniqueness result.

\begin{proposition}[Local well-posedness of propagaton dynamics]
\label{prop:rfd_local_wellposedness}
Let $g$ be a positive-definite $C^2$ metric on an open regular domain $U\subseteq\operatorname{Dom}(\mathcal V)$, and suppose that $\mathcal V\in C^{1,1}_{\mathrm{loc}}(U)$. Let $\Lambda>0$, $m_*,\nu\geq0$, and $(m_*,\nu)\ne(0,0)$. For $m_*=0$, the massless dynamical law~\eqref{eq:rfd_massless_propagaton_motion} admits a unique maximal solution for each initial configuration $q_0\in U$. For $m_*>0$, the unified dynamical law~\eqref{eq:rfd_unified_geometric_motion} admits a unique maximal solution for each initial state $(q_0,v_0)\in TU$.
\end{proposition}
\begin{proof}
In local coordinates $q^a$, the massless vector field has components $\zeta g^{ab}\partial_b\mathcal V$. For $m_*>0$, the inertial law admits the equivalent first-order formulation on $TU$,
$\dot q^a=v^a$, $\dot v^a=-\Gamma^a_{bc}v^bv^c+(\Lambda/m_*)g^{ab}\partial_b\mathcal V-(\nu/m_*)v^a$.
The assumed regularity of $g$ and $\mathcal V$ ensures that both vector fields are locally Lipschitz in their respective state spaces. The existence, uniqueness, and maximal continuation of their integral curves follow from the Picard--Lindel\"of theorem~\cite{HirschSmaleDevaney2013}. The coordinate independence of the underlying dynamical laws ensures that the local solutions define intrinsic trajectories on their respective manifolds.
\end{proof}

The regularity hypothesis on $\mathcal V$ in Proposition~\ref{prop:rfd_local_wellposedness} holds throughout $\operatorname{Dom}(\mathcal V)$ for smooth compactly supported sources $J\in C_c^\infty(\mathbb R^3)$, by Proposition~\ref{prop:rfd_smooth_rotated_convolutions}. Proposition~\ref{prop:rfd_weak_source_regularity} extends this conclusion to compactly supported H\"older-continuous sources $J\in C_c^{0,\alpha}(\mathbb R^3)$, $0<\alpha<1$, and to bounded compactly supported sources on regular configurations outside the source support $D_J$. The relevant function spaces are defined in Appendix~\ref{sec:rfd_analytical_structure}. Under the metric hypotheses of Proposition~\ref{prop:rfd_local_wellposedness}, this extension makes locally well-posed propagaton dynamics accessible to a wider class of physical systems, including exterior observations of bounded sources with abrupt spatial boundaries. The continuation of propagaton trajectories is further influenced by the geometry of the regular domain, particularly in the vicinity of zeros of the propagating partial field $\varphi_{\mathrm{pr}}$. The configuration-space energy balance~\eqref{eq:rfd_configuration_energy_identity} provides a means of characterizing the accessibility of such configurations.

\begin{proposition}[Energetic restriction on accessible configurations]
\label{prop:rfd_regular_domain_barrier}
Along any forward regular solution of the unified dynamical law, with $q(0)=q_0$, the relative evanescent strength satisfies
$\eta(q(\tau))\leq\mathcal E_{\mathrm{cfg}}(0)/\Lambda=\eta(q_0)+m_*\|\dot q(0)\|_g^2/(2\Lambda)$.
In particular, for the massless branch, $\eta(q(\tau))\leq\eta(q_0)$. If the trajectory converges to a finite configuration $q_*$ at which the partial fields are continuous and $\varphi_{\mathrm{pr}}(q_*)=0$, then $\varphi_{\mathrm{ev}}(q_*)=0$.
\end{proposition}
\begin{proof}
The configuration-space energy satisfies $\mathcal E_{\mathrm{cfg}}\geq\Lambda\eta$ by the nonnegativity of the kinetic energy. The energy balance~\eqref{eq:rfd_configuration_energy_identity} gives $\mathcal E_{\mathrm{cfg}}(\tau)\leq\mathcal E_{\mathrm{cfg}}(0)$ for forward evolution, establishing the bound. Since $e=\eta p$ on the regular domain, continuity at $q_*$ gives $0\leq e(q_*)\leq[\mathcal E_{\mathrm{cfg}}(0)/\Lambda]p(q_*)=0$, and hence $\varphi_{\mathrm{ev}}(q_*)=0$.
\end{proof}

Proposition~\ref{prop:rfd_regular_domain_barrier} confines forward trajectories with initial energy $E_0=\mathcal E_{\mathrm{cfg}}(0)$ to the accessible region $\{q\in\operatorname{Dom}(\mathcal V):\eta(q)\leq E_0/\Lambda\}$. When the propagating amplitude approaches zero while the evanescent amplitude tends to a finite nonzero value, the mechanical potential energy $-\Lambda\mathcal V$ diverges to positive infinity, producing a singular barrier to forward evolution at finite initial energy. The geometrical structure of the remaining singular configurations depends on the phase structure of the source field $J$. For a source of uniform phase, $\varphi_{\mathrm{ev}}$ is real up to a constant phase factor (Remark~\ref{rem:rfd_real_evanescent}), and its zero set is a hypersurface under the corresponding regular-value hypothesis. For general complex sources, regular zero sets of the individual complex partial fields have codimension two. Under the respective transversality hypotheses, simultaneous zeros have codimension three for uniform-phase sources and four for general complex sources. Global continuation consequently involves controlling the approach to simultaneous zeros, the behavior of trajectories at spatial infinity, and, in the inertial case, the evolution of tangent velocities.

\subsection{Intrinsic axial symmetry}
\label{sec:rfd_symmetry_structure}

\begin{figure}[!tbp]
\centering
\begingroup
  \colorlet{rfdink}{black}\colorlet{rfdplane}{blue!4!white}
  \colorlet{rfdbasis}{blue!65!black}\colorlet{rfdfaint}{black!45!white}
\begin{tikzpicture}[x=1cm,y=1cm,>=stealth, font=\small, every node/.style={text=rfdink}]
  \foreach \cx/\ang/\panel/\framename/\basisone/\basistwo in {0/15/a/{\gamma}/{\hat{x}'_1}/{\hat{x}'_2},6.65/60/b/{\gamma h}/{\hat{x}''_1}/{\hat{x}''_2}} {
    \begin{scope}[shift={(\cx,0)}]
      \filldraw[fill=rfdplane,draw=rfdfaint,line width=.45pt] (0,0) ellipse (1.65 and .68);
      \draw[rfdfaint,densely dashed] (0,-.15)--(0,-.98);
      \draw[rfdink,->,line width=1pt] (0,0)--(0,1.88) node[above] {$\hat\alpha$};
      \ifnum\ang=60\relax
        \draw[rfdbasis,->,line width=1.1pt] (0,0)--({1.44*cos(\ang)},{-.59*sin(\ang)}) node[anchor=north east,xshift=-1pt,yshift=-2pt] {$\basisone$};
      \else
        \draw[rfdbasis,->,line width=1.1pt] (0,0)--({1.44*cos(\ang)},{-.59*sin(\ang)}) node[below right] {$\basisone$};
      \fi
      \draw[rfdbasis,->,line width=1.1pt] (0,0)--({1.44*cos(\ang+90)},{-.59*sin(\ang+90)}) node[below left] {$\basistwo$};
      \fill[rfdink] (0,0) circle (1.6pt) node[above left=2pt] {$\mathbf x$};
      \node[anchor=west] at (-1.88,1.85) {(\panel)};
      \node at (0,-1.32) {$\framename$};
    \end{scope}
  }
  \draw[rfdink,->,line width=.8pt] (2.02,.52)--(4.62,.52)
    node[midway,above=3pt] {$\mathcal R_h$}
    node[midway,below=3pt] {$h\in\mathcal G_{\mathrm{ax}}$};
  \begin{scope}[shift={(6.65,0)}]
    \draw[rfdfaint,dashed,->,line width=.7pt] (0,0)--({1.44*cos(15)},{-.59*sin(15)});
    \draw[rfdfaint,dashed,->,line width=.7pt] (0,0)--({1.44*cos(105)},{-.59*sin(105)});
    \draw[rfdbasis,->,line width=.8pt,domain=15:60,samples=30,variable=\t]
       plot ({1.92*cos(\t)},{-.79*sin(\t)});
    \node[text=rfdbasis] at (1.83,-.83) {$\theta$};
  \end{scope}
\end{tikzpicture}
\endgroup
\caption{Intrinsic axial symmetry of the local spectral frame at fixed $J$ and $\mathbf x$. (a) Frame $\gamma$ with transverse basis $\hat{x}'_j=\gamma\hat{x}_j$ ($j=1,2$). (b) Frame $\gamma h$ after an axial rotation through $\theta$ (dashed arrows indicate the original transverse directions). The longitudinal axis $\hat\alpha=\gamma\hat{x}_3=(\gamma h)\hat{x}_3$ is preserved, as are the partial fields and the propagation potential on its regular domain (Proposition~\ref{prop:rfd_axial_rotation_invariance} and Corollary~\ref{prop:rfd_potential_differential_symmetry}).}
\label{fig:rfd_axial_frame_symmetry}
\end{figure}
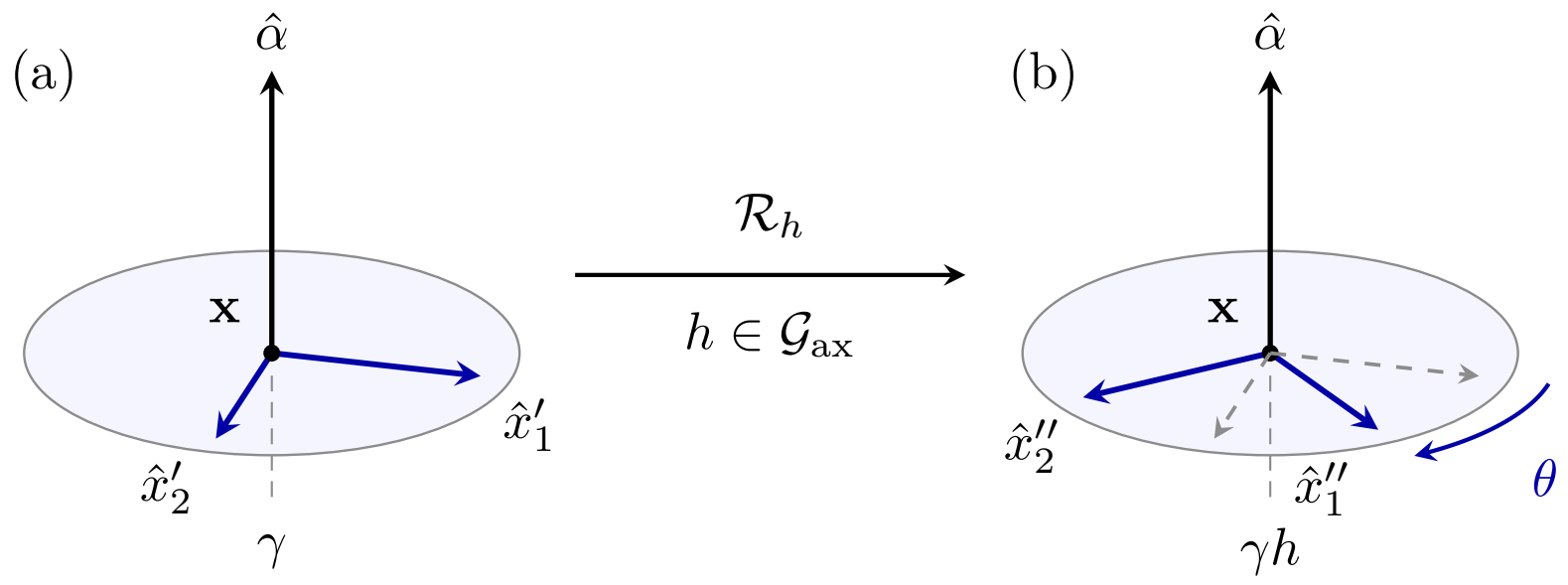

The introduction of the local spectral frame $\gamma$ into the configuration space $\mathcal M$ raises the question of how its three rotational degrees of freedom enter the spectral structure of the radiation field. The Weyl decomposition distinguishes the longitudinal direction, specified by the local spectral axis $\hat\alpha=\gamma\hat{x}_3$, from the transverse spectral plane. While changes in $\hat\alpha$ generally alter the partial fields (Appendix~\ref{app:rfd_point_source_rotation}), rotations of the transverse basis about $\hat\alpha$ leave their spectral partition invariant. This freedom constitutes an intrinsic continuous symmetry of the source--response construction, holding for arbitrary admissible spatial source distributions $J$, including sources with no rotational symmetry. To investigate its consequences for the dynamics, we introduce the axial stabilizer subgroup and its right action on $\mathcal M$,
\begin{equation}
\label{eq:rfd_axial_group_action}
\mathcal G_{\mathrm{ax}}=\{h\in\mathrm{SO}(3):h\hat{x}_3=\hat{x}_3\}\simeq\mathrm{SO}(2),\qquad
\mathcal R_h(\mathbf x,\gamma)=(\mathbf x,\gamma h).
\end{equation}
The action preserves both $\mathbf x$ and $\hat\alpha$ and rotates the transverse spectral coordinates within their plane. Figure~\ref{fig:rfd_axial_frame_symmetry} illustrates this freedom at a fixed spatial point. In particular, the propagating disk $|\mathbf k_{\mathrm t}|^2<k^2$ and the evanescent exterior $|\mathbf k_{\mathrm t}|^2>k^2$ are invariant under this rotation (Eqs.~\eqref{eq:G_pr} and~\eqref{eq:G_ev}). The corresponding invariance of the partial fields follows from the rotated kernel representations.

\begin{proposition}[Axial invariance of the partial fields]
\label{prop:rfd_axial_rotation_invariance}
For $h\in\mathcal G_{\mathrm{ax}}$ and $\iota\in\{\mathrm{pr},\mathrm{ev}\}$,
\begin{equation}
\label{eq:rfd_axial_stabilizer}
\varphi_\iota(\mathbf x,\gamma h)=\varphi_\iota(\mathbf x,\gamma).
\end{equation}
\end{proposition}
\begin{proof}
Under $\gamma\mapsto\gamma h$, the longitudinal separation $z_\gamma$ and the transverse magnitude $|\mathbf s_{\mathrm t,\gamma}|$ are preserved. Both kernels in Eqs.~\eqref{eq:rfd_rotated_propagating_kernel} and~\eqref{eq:rfd_rotated_evanescent_kernel} depend only on these quantities. Their convolutions with $J$ therefore satisfy the stated identity.
\end{proof}

Proposition~\ref{prop:rfd_axial_rotation_invariance} identifies a one-parameter family of spectrally equivalent frames at every configuration. By Definition~\ref{def:rfd_local_propagation_potential}, the propagation potential is constant on each such orbit, and its differential inherits the corresponding symmetry. This property will be central to the momentum law and the reduction of the dynamics in Section~\ref{sec:rfd_axial_momentum_reduction}.
\begin{corollary}[Axial symmetry of the potential and its differential]
\label{prop:rfd_potential_differential_symmetry}
The regular domain $\operatorname{Dom}(\mathcal V)$ is invariant under $\mathcal R_h$, and
\begin{equation}
\label{eq:rfd_potential_pullback_symmetry}
\mathcal V(\mathcal R_hq)=\mathcal V(q),\qquad
\mathrm d\mathcal V_{\mathcal R_hq}[(\mathrm d\mathcal R_h)_qv]=\mathrm d\mathcal V_q[v].
\end{equation}
In particular, $\mathrm d\mathcal V$ vanishes on the infinitesimal generator of the axial action.
\end{corollary}
\begin{proof}
Invariance of the squared partial-field amplitudes gives invariance of the regular domain and the first identity. Differentiating the scalar identity with respect to $q$ yields the second, while differentiation along the axial one-parameter subgroup yields the last assertion.
\end{proof}

The spectral equivalence of two configurations suggests investigating whether their subsequent evolution under the unified dynamical law~\eqref{eq:rfd_unified_geometric_motion} also remains equivalent. Since the metric $g$ converts the differential of $\mathcal V$ into the driving vector field and determines the inertial connection, compatibility of the geometry with the axial action supplies a sufficient condition for dynamical equivariance. This condition applies to the full metric, including possible spatial--rotational cross terms.
\begin{proposition}[Axial equivariance of the dynamical laws]
\label{prop:rfd_gradient_equivariance}
If each $\mathcal R_h$ is an isometry of $g$, i.e.,
\begin{equation}
\label{eq:rfd_metric_isometry_requirement}
g_{\mathcal R_hq}((\mathrm d\mathcal R_h)_qv,(\mathrm d\mathcal R_h)_qw)=g_q(v,w),
\end{equation}
then
\begin{equation}
\label{eq:rfd_gradient_equivariance}
\operatorname{grad}_g\mathcal V(\mathcal R_hq)
=(\mathrm d\mathcal R_h)_q\operatorname{grad}_g\mathcal V(q).
\end{equation}
Both the massless and the inertial--relaxational equations are equivariant under $\mathcal R_h$. Axially related solutions with appropriately transformed initial velocities have identical spatial projections.
\end{proposition}
\begin{proof}
Metric duality~\eqref{eq:rfd_riemannian_gradient_definition}, the isometry condition~\eqref{eq:rfd_metric_isometry_requirement}, and the differential identity~\eqref{eq:rfd_potential_pullback_symmetry} give the gradient transformation law. Since isometries preserve the Levi--Civita connection, the velocity and covariant acceleration transform by the same tangent map as the driving term in Eq.~\eqref{eq:rfd_unified_geometric_motion}. The spatial statement follows from uniqueness (Proposition~\ref{prop:rfd_local_wellposedness}) and the invariance of $\mathbf x$ under $\mathcal R_h$.
\end{proof}

The axial action describes an internal symmetry of the spectral frame at fixed source $J$ and position $\mathbf x$. A complementary transformation property is the joint Euclidean covariance of the source--response arrangement established in Proposition~\ref{prop:rfd_joint_source_observer_covariance}. Let $\mathcal T_{\mathrm R,\mathbf b}(\mathbf x,\gamma)=(\mathrm R\mathbf x+\mathbf b,\mathrm R\gamma)$, with $\mathrm R\in\mathrm{SO}(3)$ and $\mathbf b\in\mathbb R^3$, and transform the source to $J_{\mathrm R,\mathbf b}$. If the metrics for the two source problems satisfy the isometry condition $g_{J_{\mathrm R,\mathbf b}}(\mathrm d\mathcal T_{\mathrm R,\mathbf b}v,\mathrm d\mathcal T_{\mathrm R,\mathbf b}w)=g_J(v,w)$, the covariance identity~\eqref{eq:rfd_joint_covariance} yields the corresponding transformation of the gradients and the Levi--Civita connections. Consequently, a solution $q_J(\tau)$ is carried to the solution $\mathcal T_{\mathrm R,\mathbf b}q_J(\tau)$ of the transformed source problem, with correspondingly transformed initial data. The source-independent product metric $g_0$ introduced in Section~\ref{sec:rfd_geometric_dynamics} satisfies both the axial and the joint Euclidean isometry conditions. The role of metric compatibility is illustrated by the example in Appendix~\ref{app:rfd_quotient_technical}, where an axially invariant scalar and a noninvariant metric produce a gradient that varies along the axial orbit. The isometric axial action also supplies the geometrical basis for the momentum law and the reduction of the dynamics developed next.

\subsection{Axial momentum and geometric reduction}
\label{sec:rfd_axial_momentum_reduction}

The intrinsic axial symmetry established in Section~\ref{sec:rfd_symmetry_structure} gives rise to a momentum associated with the rotational degrees of freedom of the local spectral frame. For an axially invariant metric $g$, both the kinetic and potential terms of the massive Lagrangian~\eqref{eq:rfd_massive_lagrangian} are invariant under $\mathcal G_{\mathrm{ax}}$. The infinitesimal generator of the axial action is therefore a Killing vector field, and the corresponding Noether momentum has a direct mechanical interpretation. The unified dynamical law also allows us to determine how this momentum evolves when inertia and relaxation are present simultaneously.
\begin{proposition}[Evolution of the axial momentum]
\label{prop:rfd_axial_noether}
Let $h(\theta)\in\mathcal G_{\mathrm{ax}}$ be rotation through angle $\theta$ about $\hat{x}_3$, and let $Z_q=\left.\frac{\mathrm d}{\mathrm d\theta}\right|_{\theta=0}\mathcal R_{h(\theta)}(q)$ be the infinitesimal generator. If $g$ is invariant under $\mathcal G_{\mathrm{ax}}$ and $m_*>0$, then along every solution of the unified dynamical law~\eqref{eq:rfd_unified_geometric_motion}, with initial evolution parameter $\tau=0$,
\begin{equation}
\label{eq:rfd_axial_noether_integral}
\mathcal J_{\mathrm{ax}}\coloneqq m_*g_q(\dot q,Z_q),\qquad
\frac{\mathrm d\mathcal J_{\mathrm{ax}}}{\mathrm d\tau}=-\frac{\nu}{m_*}\mathcal J_{\mathrm{ax}},
\qquad \mathcal J_{\mathrm{ax}}(\tau)=\mathcal J_{\mathrm{ax}}(0)\,e^{-\nu\tau/m_*}.
\end{equation}
The momentum $\mathcal J_{\mathrm{ax}}$ is conserved for $\nu=0$, the sector $\mathcal J_{\mathrm{ax}}=0$ is invariant, and for $\nu>0$ the momentum tends to zero along every forward-global solution.
\end{proposition}
\begin{proof}
Since $Z$ is Killing, $g(\dot q,\nabla^g_{\dot q}Z)=0$. Differentiating the momentum pairing and using the unified dynamical law together with $\mathrm d\mathcal V[Z]=0$ (Corollary~\ref{prop:rfd_potential_differential_symmetry}) gives
$\mathrm d\mathcal J_{\mathrm{ax}}/\mathrm d\tau=m_*g(\nabla^g_{\dot q}\dot q,Z)=-\nu g(\dot q,Z)=-(\nu/m_*)\mathcal J_{\mathrm{ax}}$.
Integration yields~\eqref{eq:rfd_axial_noether_integral}; for $\nu=0$, this is the Noether conservation law~\cite{MarsdenRatiu1999}.
\end{proof}

\begin{figure}[!tbp]
\centering
\begingroup
  \colorlet{rfdink}{black}\colorlet{rfdgrid}{black!15!white}
  \colorlet{rfdcurve}{blue!65!black}
\begin{tikzpicture}[x=1.65cm,y=2.85cm,>=stealth,font=\small,every node/.style={text=rfdink}]
  \foreach \v in {0.25,0.5,0.75,1} {\draw[rfdgrid,line width=.35pt] (0,\v)--(4,\v);}
  \draw[rfdink,line width=.65pt] (0,1.10)--(0,0)--(4,0);
  \foreach \s in {0,1,2,3,4} {
    \draw[rfdink] (\s,0)--++(0,-.025) node[below] {$\s$};
  }
  \foreach \v/\lab in {0/0,0.25/0.25,0.5/0.5,0.75/0.75,1/1} {
    \draw[rfdink] (0,\v)--++(-.055,0) node[left] {$\lab$};
  }
  \node[rotate=90] at (-.68,.55) {$\mathcal J_{\mathrm{ax}}(\tau)/\mathcal J_{\mathrm{ax}}(0)$};
  \node at (2,-.26) {$\tau/T_0$};
  \draw[rfdink,dashed,line width=1pt] (0,1)--(4,1);
  \draw[rfdcurve,line width=1.1pt,domain=0:4,samples=81,variable=\s] plot (\s,{exp(-\s)});
  \draw[rfdink,dashed,line width=1pt] (1.75,.79)--(2.12,.79);
  \node[anchor=west] at (2.19,.79) {$\nu T_0/m_*=0$};
  \draw[rfdcurve,line width=1.1pt] (1.75,.62)--(2.12,.62);
  \node[anchor=west] at (2.19,.62) {$\nu T_0/m_*=1$};
\end{tikzpicture}
\endgroup
\caption{Axial momentum evolution implied by Proposition~\ref{prop:rfd_axial_noether}, normalized by a nonzero initial momentum $\mathcal J_{\mathrm{ax}}(0)$ and a reference evolution scale $T_0>0$. The conservative massive branch ($\nu=0$) has constant momentum; the inertial--relaxational regime follows the exact exponential $\exp[-(\nu T_0/m_*)(\tau/T_0)]$, illustrated for $\nu T_0/m_*=1$. The curves represent the momentum law on the forward interval of existence.}
\label{fig:rfd_axial_momentum_evolution}
\end{figure}
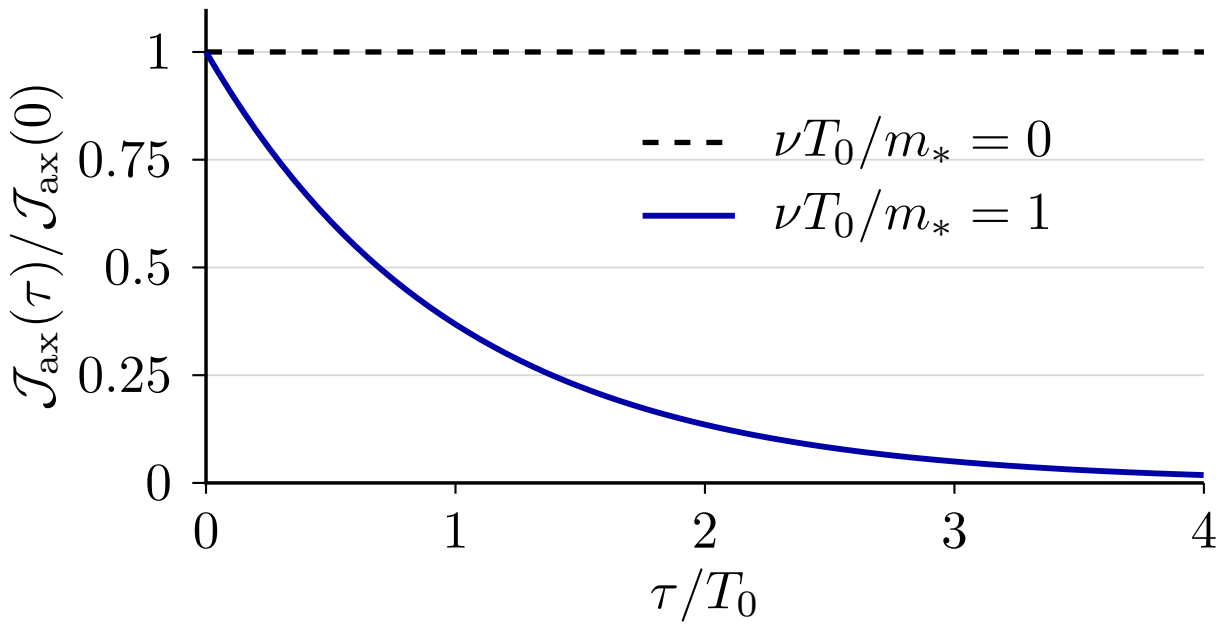

The momentum $\mathcal J_{\mathrm{ax}}$ is conjugate to rotation of the transverse spectral frame about $\hat\alpha$. For the product metric $g_0$ of Section~\ref{sec:rfd_geometric_dynamics}, define the body angular velocity $\boldsymbol\omega$ by $\gamma^{-1}\dot\gamma=\mathrm A(\boldsymbol\omega)$. The normalization of the Lie-algebra inner product then gives $\mathcal J_{\mathrm{ax}}=m_*\ell^2\boldsymbol\omega\cdot\hat{x}_3$, where $m_*\ell^2$ is the effective rotational inertia. Equation~\eqref{eq:rfd_axial_noether_integral} thus connects the spectral symmetry to an exact law for the conservation or relaxation of internal rotational momentum. Figure~\ref{fig:rfd_axial_momentum_evolution} illustrates these two regimes in dimensionless variables. The fixed axial group action preserves the kinetic term, while evolution-dependent rotations contribute an additional axial velocity to it (Appendix~\ref{app:rfd_quotient_technical}).

The axial symmetry also suggests a simplification of the dynamics: spectrally equivalent frames share the same spatial position $\mathbf x$ and local spectral axis $\hat\alpha=\gamma\hat{x}_3$. The free and proper axial action defines a principal $\mathcal G_{\mathrm{ax}}$-bundle over a five-dimensional quotient of the regular configuration space. An axially invariant metric induces a Riemannian metric on this quotient by Riemannian submersion~\cite{LeeRiemannian2018,MarsdenRatiu1999}. The axial momentum law identifies the invariant sector in which the full inertial dynamics projects to a natural mechanical system on the position--axis space.
\begin{proposition}[Conditional five-dimensional reduction]
\label{prop:rfd_five_dimensional_reduction}
Let $\mathcal U=\operatorname{Dom}(\mathcal V)$ and assume a positive-definite $C^2$ metric $g$ on $\mathcal U$ invariant under $\mathcal G_{\mathrm{ax}}$. The free, proper right group action has a smooth five-dimensional quotient $\mathcal Q=\mathcal U/\mathcal G_{\mathrm{ax}}$, identifiable with an open subset of $\mathbb R^3\times S^2$. The natural projection
\begin{equation}
\label{eq:rfd_quotient_projection}
\mathfrak p:\mathcal U\longrightarrow\mathcal Q,\qquad
\mathfrak p(\mathbf x,\gamma)=(\mathbf x,\gamma\hat{x}_3)
\end{equation}
is a Riemannian submersion for a uniquely induced quotient metric $\bar g$, and $\mathcal V=\bar{\mathcal V}\circ\mathfrak p$. Writing $\bar q=\mathfrak p(q)$, the projected equations are
\begin{enumerate}
\item for the massless branch, $\dot{\bar q}=\zeta\operatorname{grad}_{\bar g}\bar{\mathcal V}$;
\item for a solution of the unified dynamical law with $m_*>0$ and $\mathcal J_{\mathrm{ax}}=0$, $m_*\nabla^{\bar g}_{\dot{\bar q}}\dot{\bar q}+\nu\dot{\bar q}=\Lambda\operatorname{grad}_{\bar g}\bar{\mathcal V}$, which for $\nu=0$ is the conservative reduced law.
\end{enumerate}
\end{proposition}
The proof and explicit construction of the quotient metric are given in Appendix~\ref{app:rfd_quotient_technical}. The condition $\mathcal J_{\mathrm{ax}}=0$ selects horizontal velocities and is preserved by Proposition~\ref{prop:rfd_axial_noether}. For the product metric, the reduced metric is $\bar g=\mathrm d\mathbf x\cdot\mathrm d\mathbf x+\ell^2\mathrm d\hat\alpha\cdot\mathrm d\hat\alpha$. This reduction underlies the numerical investigation in Section~\ref{sec:rfd_interpretation} and Appendix~\ref{app:rfd_numerical_dynamics}: the massive trajectory begins at rest and hence lies in the zero-momentum sector, while the massless gradient is horizontal by construction. At nonzero axial momentum, the reduced inertial dynamics generally includes additional momentum-dependent terms involving the bundle geometry. Local choices of transverse frame are sections of the principal bundle, whose rotational factor $\mathrm{SO}(3)\to S^2$ has no global smooth section. The quotient therefore gives a global geometrical description of frame equivalence classes, with the full frame recoverable by horizontal lifting in the zero-momentum sector once an initial frame is specified.

The free-space spectral kernels also possess the discrete symmetry $\hat\alpha\mapsto-\hat\alpha$, through their dependence on $|z_\gamma|$ and the transverse separation. At the level of the partial fields and the propagation potential this permits a further identification with an open subset of $\mathbb R^3\times\mathbb{RP}^2$, where $\mathbb{RP}^2$ is the real projective plane. A corresponding dynamical reduction requires compatibility of the quotient metric $\bar g$ with axis reversal. More broadly, the emergence of a classical internal rotational momentum from the spectral structure of a scalar radiation field suggests investigating its relationship with internal angular-momentum degrees of freedom, including spin, in a future quantized formulation of propagaton dynamics.

\section{Physical Interpretation of Propagaton Dynamics}
\label{sec:rfd_interpretation}

A broader motivation for the present theory is to investigate whether motion can be understood as a fundamental phenomenon arising from relationships between fields. Starting with a prescribed source field $J$ and its radiation field $\varphi$, the spectral construction of Section~\ref{sec:rfd_source_response_potential}, together with the geometrical and dynamical principles of Section~\ref{sec:rfd_geometric_dynamics}, leads to a point-like dynamical entity in the extended configuration space $\mathcal M$. The propagaton thus represents an elementary form of motion associated with the generation of propagating waves relative to non-propagating waves within the proposed theory. Its trajectory $q(\tau)=(\mathbf x(\tau),\gamma(\tau))$ describes the evolution of both the spatial position and the local spectral frame through which the partial fields $\varphi_{\mathrm{pr}}$ and $\varphi_{\mathrm{ev}}$ are resolved. In the massless branch, this evolution is characterized by a nonincreasing relative evanescent strength $\eta=-\mathcal V$ (Eq.~\eqref{eq:rfd_massless_lyapunov_identity}), with the relative suppression of coherent interference as $\mathcal V$ approaches zero quantified by Proposition~\ref{prop:rfd_relative_interference_control}. In the conservative massive branch, the configuration-space energy balance~\eqref{eq:rfd_configuration_energy_identity} allows an exchange between kinetic energy and potential energy $-\Lambda\mathcal V$, giving rise to a richer evolution of the local spectral composition. The possibility of constructing such dynamical entities from a field-based starting point motivates a broader research programme in which propagatons may provide a foundation for other forms of motion, including those associated with matter.

A question of immediate physical interest is how propagaton motion relates to the transport of energy in the radiation field. The scalar-field energy flux $\mathbf S=-(\partial_t\varphi)\nabla\varphi$ is obtained from the canonical energy balance for the real time-dependent field (Appendix~\ref{app:pure_scalar}), whereas the projected propagaton trajectory $\mathbf x(\tau)$ arises from the variation of the propagation potential $\mathcal V$ on $\mathcal M$ and the unified dynamical law~\eqref{eq:rfd_unified_geometric_motion}. The distinction becomes particularly transparent when the total radiation field possesses a high degree of spatial symmetry while its partial fields retain a dependence on the local spectral axis $\hat\alpha=\gamma\hat{x}_3$. For an outgoing point source, at a fixed radial distance $r=\lambda/4$, the propagation potential takes the values $-1/2$ and $0$ for spectral axes parallel and perpendicular to the radial direction, respectively (Proposition~\ref{prop:rfd_point_orientation_contrast}). Thus even within a spherically symmetric radiation field, the dependence of $\mathcal V$ on the spectral frame supplies an additional structure for the generation of motion. The radial time-averaged energy flux of the point-source field provides a natural analytical reference against which to examine the spatial trajectories generated by both propagaton laws.

\begin{figure}[!tbp]
\centering
\begingroup
  \colorlet{rfdink}{black}\colorlet{rfdfaint}{black!50!white}
  \colorlet{rfdspectral}{blue!65!black}\colorlet{rfdvelocity}{orange!70!black}
  \colorlet{rfdflux}{green!45!black}
\begin{tikzpicture}[x=1cm,y=1cm,>=stealth,font=\small,every node/.style={text=rfdink}]
  \coordinate (O) at (0,0);
  \coordinate (P) at (3.7,0);
  \draw[rfdfaint,densely dashed,line width=.65pt] (O)--(P)
    node[midway,below=3pt] {$r$};
  \fill[rfdink] (O) circle (1.8pt) node[below=5pt] {Point source};
  \begin{scope}[shift={(P)}]
    \draw[rfdflux,->,line width=1.05pt] (0,0)--(3.45,0)
       node[right=3pt,text=rfdflux] {$\hat r\parallel\langle\mathbf S\rangle$};
    \draw[rfdvelocity,->,line width=1.05pt] (0,0)--(34:3.38)
       node[above right=1pt,text=rfdvelocity] {$\hat v_\tau$};
    \draw[rfdspectral,->,line width=1.05pt] (0,0)--(76:3.10)
       node[above=2pt,text=rfdspectral] {$\hat\alpha$};
    \draw[rfdink,line width=.65pt] (0:1.02) arc[start angle=0,end angle=34,radius=1.02];
    \node[anchor=west] at (17:1.16) {$\vartheta_{\mathrm F}$};
    \draw[rfdink,line width=.65pt] (34:1.53) arc[start angle=34,end angle=76,radius=1.53];
    \node at (55:1.98) {$\arccos\mu$};
    \fill[rfdink] (0,0) circle (1.8pt) node[below=5pt] {$\mathbf x$};
  \end{scope}
\end{tikzpicture}
\endgroup
\caption{Local geometry of propagaton motion relative to the outward time-averaged energy flux of a point-source field. At the observation point $\mathbf x$, the alignment cosine $\mu=\hat\alpha\cdot\hat v_\tau$ characterizes the direction of projected motion relative to the local spectral axis, whereas $\vartheta_{\mathrm F}=\arccos(\hat v_\tau\cdot\hat r)$ measures its angular departure from radial energy transport. The three directions are drawn coplanar for illustration.}
\label{fig:rfd_local_motion_geometry}
\end{figure}
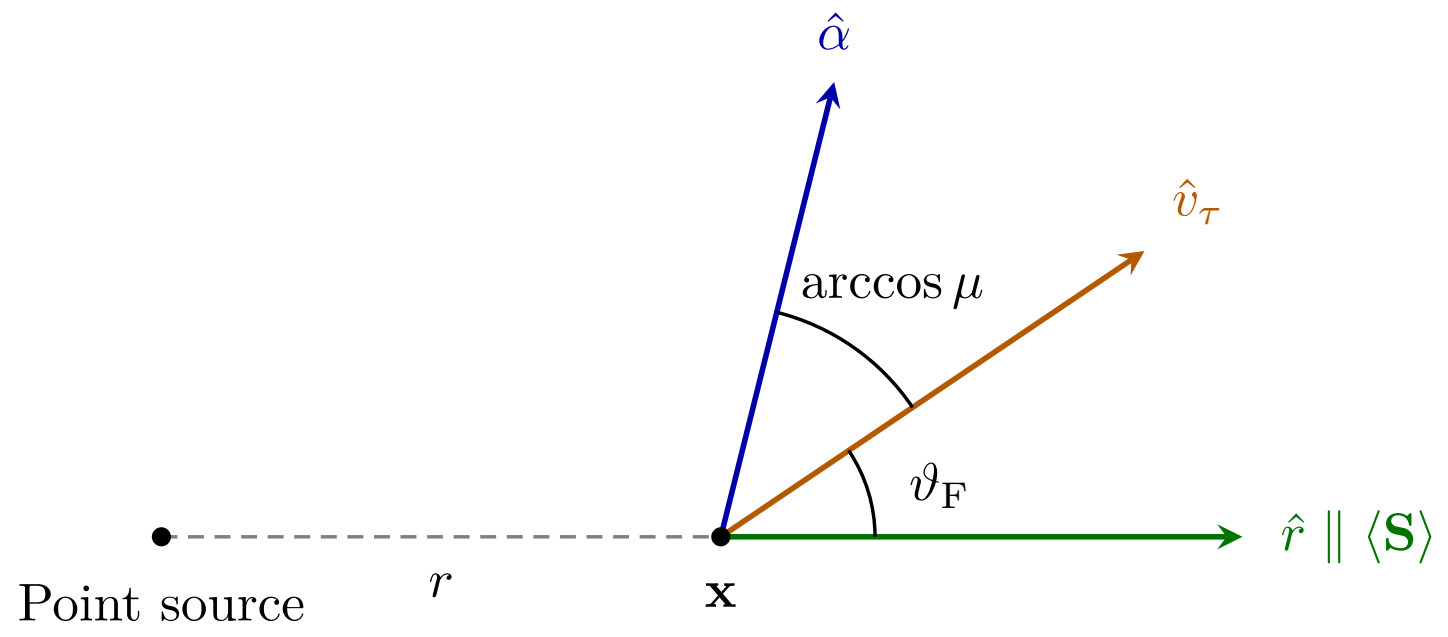

For a unit point source located at the origin, the radiation field is given by $\varphi(r)=\exp(\im kr)/(4\pi r)$, where $r=|\mathbf x|>0$. Using the canonical energy balance developed in Appendix~\ref{app:pure_scalar}, the time-averaged energy flux associated with this field is obtained as
\begin{equation}
\label{eq:rfd_point_flux}
\langle\mathbf S\rangle=\frac{\omega}{2}\operatorname{Im}(\varphi^*\nabla\varphi)
=\frac{\omega k}{32\pi^2r^2}\hat r.
\end{equation}
Here $\hat r=\mathbf x/r$, so the streamlines of $\langle\mathbf S\rangle$ are outward radial rays, providing an analytical reference for the investigation of propagaton motion. To characterize the relationship between the projected propagaton velocity $\mathbf v_\tau$, the local spectral axis $\hat\alpha$, and the direction of energy transport $\hat r$, we employ two complementary geometrical quantities: the alignment cosine $\mu=\hat\alpha\cdot\hat v_\tau$ introduced in Section~\ref{sec:rfd_unified_dynamics} and the flux angle $\vartheta_{\mathrm F}=\arccos(\hat v_\tau\cdot\hat r)$, evaluated wherever the projected velocity is nonzero. As illustrated in Fig.~\ref{fig:rfd_local_motion_geometry}, the former characterizes the alignment of propagaton motion with the direction along which the spectral decomposition is performed, while the latter measures its angular departure from conventional energy transport. These quantities provide the basis for the comparative investigation of massless and massive propagaton trajectories presented below.

\begin{figure}[!t]
\centering
\includegraphics[width=0.99\linewidth]{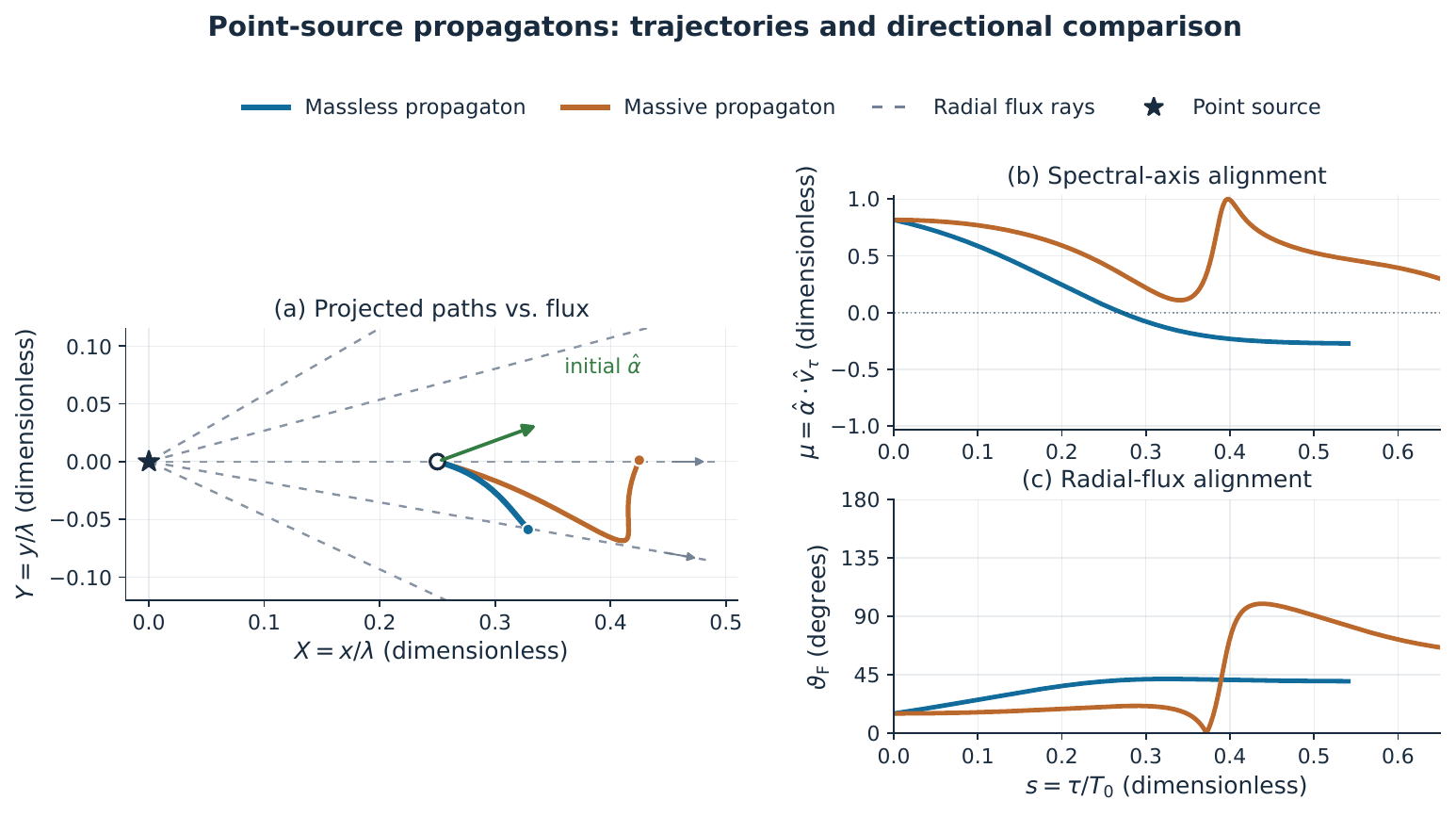}
\caption{Spatial trajectories and geometrical alignments of massless and massive propagatons in the radiation field of a point source, computed with the parameters of Table~\ref{tab:rfd_parameters}. (a) Projected trajectories in wavelength-normalized coordinates $X=x/\lambda$ and $Y=y/\lambda$, with dashed rays representing radial energy-flux streamlines. (b) Alignment cosine $\mu$ between the projected velocity and the local spectral axis. (c) Angular departure $\vartheta_{\mathrm F}$ from the radial energy flux, in degrees. The evolution parameter $s=\tau/T_0$ is dimensionless. Angular diagnostics are evaluated above the projected-speed threshold specified in Appendix~\ref{app:rfd_numerical_dynamics}.}
\label{fig:rfd_pointsource_dynamical_comparison}
\end{figure}

For the numerical investigation, we adopt the product metric $g_0$ introduced in Section~\ref{sec:rfd_geometric_dynamics}, with the characteristic rotational length $\ell=k^{-1}=\lambda/(2\pi)$, relating the geometrical weighting of spectral rotations to a characteristic length scale of the radiation field. The axial invariance of $g_0$ permits the reduction of the massless and zero-axial-momentum inertial dynamics to the regular part of the position--axis space $\mathbb R^3\times S^2$ (Proposition~\ref{prop:rfd_five_dimensional_reduction}). For the point-source field, reflection symmetry further permits both forms of motion to be investigated within an invariant planar sector (Appendix~\ref{app:rfd_numerical_dynamics}). Introducing $X=x/\lambda$, $Y=y/\lambda$, $s=\tau/T_0$ for an arbitrary evolution scale $T_0$, and the spectral axis $\hat\alpha=(\cos\psi,\sin\psi,0)$, where $\psi$ is measured from the positive Cartesian $x$-axis, we obtain the dimensionless configuration $Q=(X,Y,\psi)$ with metric $\widetilde g=\mathrm dX^2+\mathrm dY^2+(4\pi^2)^{-1}\mathrm d\psi^2$ ($\psi$ in radians). The unified dynamical law~\eqref{eq:rfd_unified_geometric_motion} becomes $\bar m\nabla^{\widetilde g}_{Q'}Q'+\bar\nu Q'=\operatorname{grad}_{\widetilde g}\mathcal V(Q)$, where the prime denotes differentiation with respect to $s$, and $\bar m$ and $\bar\nu$ are the dimensionless coefficients introduced in Section~\ref{sec:rfd_massive_inertial_branch} with $L_0=\lambda$. For the massless branch the normalized mobility is $\bar\zeta=\zeta T_0/\lambda^2=1/\bar\nu$. Table~\ref{tab:rfd_parameters} summarizes the geometrical parameters, dynamical coefficients and initial conditions. Both trajectories originate from the same spatial position and spectral axis, allowing us to examine the influence of the different dynamical laws on the subsequent evolution. The propagation potential and its derivatives are evaluated using the finite spectral representation~\eqref{eq:rfd_point_pr_finite} of Appendix~\ref{app:rfd_point_source_rotation}, while the numerical integration and convergence checks are detailed in Appendix~\ref{app:rfd_numerical_dynamics}. The resulting spatial trajectories and their alignment with the local spectral axis and radial energy flux are presented in Fig.~\ref{fig:rfd_pointsource_dynamical_comparison}.

\begin{table}[htbp]
\centering\small
\caption{Geometrical parameters, dynamical coefficients and initial conditions for the point-source trajectories. The angular coordinate $\psi$ is expressed in degrees in the table.}
\label{tab:rfd_parameters}
\begin{tabular}{@{}p{0.46\linewidth}p{0.24\linewidth}p{0.24\linewidth}@{}}
\toprule
Quantity & Massless branch & Massive branch\\\midrule
Normalized rotational length $\ell/\lambda$ & $1/(2\pi)$ & $1/(2\pi)$\\
Dimensionless inertia $\bar m$ & $0$ & $1$\\
Dimensionless relaxation $\bar\nu$ & $10$ & $0$\\
Dimensionless mobility $\bar\zeta$ & $0.1$ & ---\\
Initial position $(X_0,Y_0)$ & $(0.25,0)$ & $(0.25,0)$\\
Initial spectral angle $\psi_0$ & $20^\circ$ & $20^\circ$\\
Initial tangent velocity $Q'(0)$ & From Eq.~\eqref{eq:rfd_massless_propagaton_motion} & $\mathbf 0$\\
Evolution interval $s$ & $[0,0.65]$ & $[0,0.65]$\\\bottomrule
\end{tabular}
\end{table}

Consider first the evolution of the massless propagaton in Fig.~\ref{fig:rfd_pointsource_dynamical_comparison}. Starting from the normalized radial distance $r/\lambda=0.25$, with the local spectral axis inclined at $20^\circ$ to the radial direction, the propagaton evolves through simultaneous spatial displacement and rotation of its spectral frame. The projected trajectory in Fig.~\ref{fig:rfd_pointsource_dynamical_comparison}a departs from the radial energy-flux streamline passing through its initial position, with the flux angle $\vartheta_{\mathrm F}$ increasing from approximately $15^\circ$ to a maximum of $42^\circ$ over the portions with sufficient projected speed (Fig.~\ref{fig:rfd_pointsource_dynamical_comparison}c). Meanwhile, the alignment cosine $\mu$ changes sign during the evolution (Fig.~\ref{fig:rfd_pointsource_dynamical_comparison}b), indicating a transition between acute and obtuse angles formed by the projected velocity direction $\hat v_\tau$ and the local spectral axis $\hat\alpha$. These geometrical changes reflect the joint spatial and rotational dependence of the propagation potential $\mathcal V$. The resulting motion thus exhibits a richer directional structure than the radial transport of energy in the underlying spherically symmetric radiation field.

The spectral evolution accompanying this motion is characterized by a monotonic increase of $\mathcal V$ from approximately $-0.462$ to $-1.6\times10^{-6}$ over the computed interval, in accordance with Eq.~\eqref{eq:rfd_massless_lyapunov_identity} and the numerical consistency results in Appendix~\ref{app:rfd_numerical_dynamics}. At the final configuration, the normalized radial distance is $r/\lambda\approx0.334$, and the inclination $\theta=\arccos|\hat\alpha\cdot\hat r|$ of the unoriented spectral axis relative to the radial direction is approximately $66^\circ$. The corresponding normalized evanescent amplitude is $|4\pi rG_{\mathrm{ev}}|\approx1.3\times10^{-3}$, placing the propagaton in the vicinity of the evanescent-free configurations investigated in Remark~\ref{rem:rfd_real_evanescent}. At regular differentiable points of this equilibrium set, the propagation potential attains its maximum $\mathcal V=0$ and its metric gradient vanishes. The progressive suppression of the relative evanescent contribution along the computed trajectory is accompanied by a substantial slowing of the motion as it approaches this equilibrium structure. This example illustrates how the spectral composition of a radiation field supplies both the driving potential for motion and the geometrical structure of configurations at which such motion becomes stationary.

The conservative massive propagaton exhibits a qualitatively different form of motion, governed by the exchange between configuration-space kinetic energy and potential energy $-\Lambda\mathcal V$ established in Section~\ref{sec:rfd_massive_inertial_branch}. Although both propagatons originate from the same spatial position and local spectral axis, the massive trajectory exhibits more pronounced variations in its direction of motion (Fig.~\ref{fig:rfd_pointsource_dynamical_comparison}a). A particularly interesting feature occurs over the interval $0.37\lesssim s\lesssim0.44$, during which the flux angle $\vartheta_{\mathrm F}$ increases from nearly $0^\circ$ to approximately $100^\circ$ (Fig.~\ref{fig:rfd_pointsource_dynamical_comparison}c). The projected velocity consequently acquires an inward radial component while the energy flux of the underlying radiation field remains directed radially outward. This directional transition takes place at appreciable projected speeds, with a minimum normalized speed of approximately $0.22$ over the turning region. The simultaneous evolution of the local spectral axis, reflected in the variations of the alignment cosine $\mu$ in Fig.~\ref{fig:rfd_pointsource_dynamical_comparison}b, illustrates the dynamical coupling between spatial and spectral rotational motion induced by the propagation potential.

The spectral evolution of the massive propagaton further reveals the significance of inertia. The propagation potential approaches its maximum $\mathcal V=0$ twice along the computed trajectory, at $s\approx0.31$ and $s\approx0.44$ (Fig.~\ref{fig:rfd_pointsource_numerical_diagnostics}a). From the configuration-space energy conservation law~\eqref{eq:rfd_configuration_energy_identity}, we have $\|\dot q\|_g^2=2(\mathcal E_{\mathrm{cfg}}+\Lambda\mathcal V)/m_*$. Thus, for a conservative massive propagaton with positive configuration-space energy, a regular evanescent-free configuration, when reached, corresponds to the maximum geometrical speed permitted by its energy. This speed includes both spatial and rotational motion. In the massless branch, by contrast, regular differentiable evanescent-free configurations constitute stationary states. The inertia of the massive propagaton allows its motion to continue through the vicinity of such configurations and subsequently explore regions of greater relative evanescent strength. The computed trajectory exhibits this exchange between kinetic and potential energies, with a peak-to-peak variation of $\mathcal E_{\mathrm{cfg}}/\Lambda$ of approximately $6\times10^{-15}$ over the integration interval (Appendix~\ref{app:rfd_numerical_dynamics}).

The geometrical character of both trajectories also depends on the relative weighting of translational and rotational motion through the length parameter $\ell$. Varying $\ell$ by $25\%$ about its reference value produces changes in the spatial trajectories and their final spectral orientations while preserving their nonradial character in the investigated examples. The respective massless endpoints for the multipliers $0.75$ and $1.25$ are $(X,Y)=(0.310,-0.041)$ and $(0.343,-0.074)$, with monotone growth of $\mathcal V$ in both cases. These observations illustrate how the spectral structure of the radiation field and the geometry of the extended configuration space jointly determine propagaton motion.

The possibility of associating an elementary dynamical entity with the spectral structure of a field suggests a broader perspective on the nature of motion. An interesting connection in this respect is provided by de Broglie--Bohm theory~\cite{Bohm1952Part_I,Bohm1952Part_II,BohmHiley1982}, in which the motion of a particle is guided by an underlying quantum wavefunction. In the standard nonrelativistic formulation, the guidance velocity is obtained from the probability current and the associated quantum probability density~\cite{durr1992quantum_equilibrium,deotto1998bohmian}. The dynamical structure developed here originates instead in the relative strengths of the propagating and evanescent partial fields and their dependence on the local spectral frame. The propagaton acquires its spatial and rotational degrees of freedom through this spectral construction, with its evolution subsequently governed by the geometrical dynamical principles introduced in Section~\ref{sec:rfd_geometric_dynamics}. Furthermore, the emergence of an internal rotational momentum (Proposition~\ref{prop:rfd_axial_noether}) illustrates how the extended configuration space accommodates dynamical properties beyond those associated with spatial displacement alone. These observations motivate the investigation of propagatons as possible elementary constituents of motion, from which more familiar forms of dynamics, including those associated with material particles, might emerge.

The development of this perspective calls for extending the source--response construction to fields with general temporal dependence. A complementary direction is the determination of the metric $g$ and the dynamical coefficients $m_*$, $\nu$, and $\Lambda$ from the physical properties of the interacting fields. The dependence of propagaton trajectories on the geometry of $\mathcal M$, illustrated by the metric-sensitivity results above, suggests that this geometry may itself encode significant physical information concerning the source--response relation. The extension to vector fields and the incorporation of reciprocal dynamical interactions between source and radiation fields would considerably enlarge the scope of the theory. These developments provide a basis for investigating how propagaton dynamics might give rise to effective laws of motion in physical systems.

\section{Conclusion}
\label{sec:rfd_conclusion}

In this paper, we have developed a relational theory of motion originating in the spectral structure of source-generated fields. The central idea is to associate motion with the generation of propagating waves relative to non-propagating waves, taking the source--response relation as the physical starting point. The dependence of the propagating and evanescent partial fields, $\varphi_{\mathrm{pr}}$ and $\varphi_{\mathrm{ev}}$, on the local spectral axis leads to an extended configuration space $\mathcal M=\mathbb R^3\times\mathrm{SO}(3)$, on which their relative squared amplitudes define the propagation potential $\mathcal V$. In conjunction with a Riemannian metric $g$, this potential supplies the driving term of the unified dynamical law~\eqref{eq:rfd_unified_geometric_motion} governing the evolution of a point-like configuration, termed a propagaton. The resulting formulation provides a geometrical framework for investigating elementary forms of motion from a purely field-based starting point. Its massless branch describes gradient-driven evolution toward increasing relative propagating purity, while the massive branch incorporates inertia and, in the conservative case, admits the Lagrangian and Hamiltonian formulations of geometric mechanics (Section~\ref{sec:rfd_geometric_dynamics}).

The mathematical investigation of this theory has revealed several structural properties originating in the spectral construction. The joint Euclidean covariance of the partial fields and the propagation potential establishes the relational character of their dependence on the source, spatial position, and local spectral frame (Proposition~\ref{prop:rfd_joint_source_observer_covariance}). The analytical regularity results considerably broaden the class of admissible source distributions, extending the theory from smooth to H\"older-continuous sources of compact support and, in exterior observation regions, to general bounded sources of compact support (Propositions~\ref{prop:rfd_smooth_rotated_convolutions} and~\ref{prop:rfd_weak_source_regularity}). These results provide sufficient regularity of the propagation potential for local existence and uniqueness under the metric hypotheses of Proposition~\ref{prop:rfd_local_wellposedness}, while the configuration-space energy imposes restrictions on the accessibility of singular configurations (Proposition~\ref{prop:rfd_regular_domain_barrier}). Of particular significance is the intrinsic axial symmetry of the spectral partition, which holds for general admissible source distributions (Proposition~\ref{prop:rfd_axial_rotation_invariance}). For axially invariant metrics and $m_*>0$, this symmetry gives rise to an internal rotational momentum $\mathcal J_{\mathrm{ax}}$, conserved in the conservative massive branch and exponentially relaxed when $\nu>0$ (Proposition~\ref{prop:rfd_axial_noether}). The associated quotient geometry permits the reduction of the massless dynamics and the inertial dynamics with vanishing axial momentum to a five-dimensional position--axis configuration space (Proposition~\ref{prop:rfd_five_dimensional_reduction}). The emergence of this internal symmetry and its dynamical consequences illustrates the richness of the geometrical structure associated with the spectral decomposition of even a scalar radiation field.

The physical significance of the theory has been investigated through the radiation field of a point source. Despite the spherical symmetry of the total field, its propagation potential exhibits a nontrivial dependence on the local spectral axis (Proposition~\ref{prop:rfd_point_orientation_contrast}), giving rise to propagaton trajectories that depart from the radial streamlines of conventional field-energy transport. The numerical investigations in Section~\ref{sec:rfd_interpretation} further reveal the different dynamical roles of configurations possessing purely propagating spectral content. At regular differentiable configurations where $\varphi_{\mathrm{ev}}=0$, the massless propagaton is stationary, whereas a conservative massive propagaton with positive configuration-space energy attains its maximum energetically permitted geometrical speed upon reaching such a configuration. The spectral structure of the radiation field thus supplies both the driving potential for motion and the geometrical structure of its equilibria, with inertia determining how the propagaton evolves through this structure. These results demonstrate how distinct forms of elementary motion can be associated with the same underlying radiation field.

The broader perspective motivating Propagaton Theory is the possibility that motion in nature possesses a fundamental field-relational origin. From this perspective, propagatons may be regarded as candidates for elementary constituents of motion, through which more familiar forms of dynamics, including those associated with matter, might eventually be understood. The present monochromatic formulation provides a mathematical foundation for pursuing this possibility. Its extension to general time-dependent fields, the physical determination of configuration-space geometry, and the investigation of possible connections with quantum theories of motion constitute important directions for further research. Complementary analytical developments concern the global continuation of propagaton trajectories and the geometrical structure of simultaneous zeros of the partial fields. A systematic numerical investigation of propagaton dynamics generated by spatially extended sources will be undertaken in a second paper.

\appendix
\section{Pure Propagating Modes and Scalar Field Identities}
\label{app:pure_scalar}
A $C^1$ field $\Phi(\mathbf x,t)$ undergoes rigid translation with velocity $v\hat n$ if and only if $(\partial_t+v\hat n\cdot\nabla)\Phi=0$. While rigid translation permits arbitrary transverse structure, a \emph{pure longitudinal mode} has the form $\Phi(\mathbf x,t)=F(\hat n\cdot\mathbf x-vt)$. Its monochromatic structure is characterized by the following proposition, invoked in Section~\ref{sec:introduction}.
\begin{proposition}[Pure monochromatic propagation]
\label{thm:monochromatic_pure_mode}
Let $\Phi(\mathbf x,t)=F(\hat n\cdot\mathbf x-vt)$ be a complex-valued $C^1$ pure longitudinal mode with $v>0$. If $\Phi(\mathbf x_0,t)=a_0\exp(-\im\omega t)$ for all $t$, with $a_0\ne0$ and $\omega>0$, then $\Phi(\mathbf x,t)=\Phi_0\exp[\im(\mathbf k\cdot\mathbf x-\omega t)]$, where $\mathbf k=(\omega/v)\hat n$ and $\Phi_0\ne0$. Conversely, every such plane wave is a pure longitudinal monochromatic mode.
\end{proposition}
\begin{proof}
Setting $z_0=\hat n\cdot\mathbf x_0$ gives $F(z_0-vt)=a_0\exp(-\im\omega t)$. Since $s=z_0-vt$ ranges over $\mathbb R$, we obtain $F(s)=a_0\exp[\im(\omega/v)(s-z_0)]$. Substituting $s=\hat n\cdot\mathbf x-vt$ establishes the result, and the converse follows directly.
\end{proof}

The sourced scalar wave equation~\eqref{eq:rfd_sourced_wave_equation} follows from the Lagrangian density $\mathcal L=\tfrac12\partial_\mu\varphi\,\partial^\mu\varphi+J\varphi$, with $x^0=ct$ and Minkowski metric $g^{\mathrm M}_{\mu\nu}=\operatorname{diag}(1,-1,-1,-1)$~\cite{nastase2019classical_field_theory,rfdLandauLifshitz2000}. Stationarity of the action $S_{\mathcal O}[\varphi;J]=\int_{\mathcal O}\mathrm d^4x\,\mathcal L$ on a bounded spacetime region $\mathcal O$ under compactly supported field variations, with $J$ prescribed, yields $\Box\varphi=J$, where $\Box=\partial_\mu\partial^\mu$. For the real time-dependent field, the energy density is $\mathcal E\coloneqq\tfrac12[c^{-2}(\partial_t\varphi)^2+|\nabla\varphi|^2]$ and the energy flux is $\mathbf S\coloneqq-(\partial_t\varphi)\nabla\varphi$. Multiplication of Eq.~\eqref{eq:rfd_sourced_wave_equation} by $\partial_t\varphi$ gives the local balance $\partial_t\mathcal E+\nabla\cdot\mathbf S=J\partial_t\varphi$, with the source term representing energy exchange with the radiation field. For the monochromatic convention of Section~\ref{sec:rfd_source_response_potential}, the time-averaged flux is $\langle\mathbf S\rangle=(\omega/2)\operatorname{Im}(\varphi^*\nabla\varphi)$, where $\varphi$ denotes the complex spatial amplitude, as used in Eq.~\eqref{eq:rfd_point_flux}. For a regular localized source $J(\mathbf x,t)$, the free-space retarded solution is
\begin{equation}
\label{eq:rfd_retarded_scalar_response}
\varphi_{\mathrm{ret}}(\mathbf x,t)=\int_{\mathbb R^3}\mathrm d^3y\,\frac{J(\mathbf y,t-|\mathbf x-\mathbf y|/c)}{4\pi|\mathbf x-\mathbf y|}.
\end{equation}
Under the same monochromatic convention, this solution reduces to the outgoing Helmholtz representation~\eqref{eq:rfd_total_outgoing_response}, selected by the Sommerfeld condition $\lim_{r\to\infty}r(\partial_r\varphi-\im k\varphi)=0$, uniformly in the observation direction ($r=|\mathbf x|$)~\cite{colton2019inverse,nedelec2001acoustic}. Its locally integrable Green kernel satisfies $(\nabla^2+k^2)G_k^+=-\delta^{(3)}$ distributionally, providing the starting point for the Weyl decomposition in Appendix~\ref{app:weyl}.
\section{Fixed-Frame Weyl Decomposition}\label{app:weyl}
\label{sec:rfd_weyl_fixed_frame}

The Weyl representation of the outgoing Helmholtz Green function resolves its dependence on a selected normal direction into a transverse Fourier spectrum~\cite{Weyl1919Propagation,morse1953methods,clemmow1996the,hansen1999plane_wave,Arnoldus_Foley_2002}. In a fixed positively oriented orthonormal Cartesian basis $\{\hat{x}_1,\hat{x}_2,\hat{x}_3\}$, write $\mathbf x=\mathbf x_{\mathrm t}+z\hat{x}_3$, with $\mathbf x_{\mathrm t}=x^1\hat{x}_1+x^2\hat{x}_2$ and $z=x^3$. The transverse wavevector is $\mathbf k_{\mathrm t}=k_1\hat{x}_1+k_2\hat{x}_2$, with $\kappa\coloneqq|\mathbf k_{\mathrm t}|$. The Fourier pair is $\widehat G(\mathbf k_{\mathrm t},z)=\int_{\mathbb R^2}\mathrm d^2x_{\mathrm t}\,G_k^+(\mathbf x)e^{-\im\mathbf k_{\mathrm t}\cdot\mathbf x_{\mathrm t}}$ and $G_k^+(\mathbf x)=(2\pi)^{-2}\int_{\mathbb R^2}\mathrm d^2k_{\mathrm t}\,\widehat G(\mathbf k_{\mathrm t},z)e^{\im\mathbf k_{\mathrm t}\cdot\mathbf x_{\mathrm t}}$, understood in the tempered-distribution sense, where $\mathrm d^2x_{\mathrm t}=\mathrm dx^1\,\mathrm dx^2$ and $\mathrm d^2k_{\mathrm t}=\mathrm dk_1\,\mathrm dk_2$. Transforming the Green identity in Appendix~\ref{app:pure_scalar} yields $[\mathrm d^2/\mathrm dz^2+k^2-\kappa^2]\widehat G=-\delta(z)$. The outgoing longitudinal wavenumber $k_{\mathrm l}=\beta(\kappa)$ is given by
\begin{equation}
\label{eq:k3_branch}
\beta(\kappa)=\sqrt{k^2-\kappa^2+\im 0}
=\begin{cases}
\sqrt{k^2-\kappa^2},&0\leq\kappa<k,\\[2pt]
\im\sqrt{\kappa^2-k^2},&\kappa>k,
\end{cases}
\end{equation}
where $\im0$ denotes the outgoing limiting-absorption prescription. The grazing circle $\kappa=k$ has zero planar measure and is understood by the corresponding integrable limiting value.

\begin{proposition}[Outgoing Weyl representation]
\label{prop:rfd_weyl_green}
For $k>0$, the outgoing fundamental solution $G_k^+(\mathbf x)=e^{\im k|\mathbf x|}/(4\pi|\mathbf x|)$ admits, for $z\ne0$, the angular-spectrum representation
\begin{equation}
\label{eq:weyl_expansion}
G_k^+(\mathbf x)=\frac{\im}{2(2\pi)^2}
\int_{\mathbb R^2}\mathrm dk_1\,\mathrm dk_2\,
\frac{\exp\!\left[\im\mathbf k_{\mathrm t}\cdot\mathbf x_{\mathrm t}
+\im\beta(\kappa)|z|\right]}{\beta(\kappa)}.
\end{equation}
At $z=0$ the representation has a distributional boundary value, selected by limiting absorption, away from the coincident-source singularity.
\end{proposition}
For a derivation, see~\cite{Weyl1919Propagation,rfdFelsenMarcuvitz1994}. For $0\leq\kappa<k$, $\beta_{\mathrm p}(\kappa)\coloneqq\sqrt{k^2-\kappa^2}$ gives an oscillatory normal factor $e^{\im\beta_{\mathrm p}|z|}$; for $\kappa>k$, $\varkappa(\kappa)\coloneqq\sqrt{\kappa^2-k^2}$ gives the decaying factor $e^{-\varkappa|z|}$, with tangential oscillations and energy transport still possible~\cite{clemmow1996the,hansen1999plane_wave,Arnoldus_Foley_2002}. Thus the propagating and evanescent spectral domains are the disk $\kappa<k$ and its exterior, respectively, giving
\begin{equation}
\label{eq:rfd_green_split}
G_k^+(\mathbf x)=G_{\mathrm{pr}}(\mathbf x)+G_{\mathrm{ev}}(\mathbf x),
\end{equation}
where the \emph{two-dimensional} spectral expressions are
\begin{align}
G_{\mathrm{pr}}(\mathbf x)
&=\frac{\im}{2(2\pi)^2}\int_{\kappa<k}\mathrm dk_1\,\mathrm dk_2\,
\frac{e^{\im\mathbf k_{\mathrm t}\cdot\mathbf x_{\mathrm t}+\im\beta_{\mathrm p}(\kappa)|z|}}
{\beta_{\mathrm p}(\kappa)},
\label{eq:G_pr}\\
G_{\mathrm{ev}}(\mathbf x)
&=\frac{1}{2(2\pi)^2}\int_{\kappa>k}\mathrm dk_1\,\mathrm dk_2\,
\frac{e^{\im\mathbf k_{\mathrm t}\cdot\mathbf x_{\mathrm t}-\varkappa(\kappa)|z|}}
{\varkappa(\kappa)}.
\label{eq:G_ev}
\end{align}
The azimuthal integration in the transverse spectral plane uses the Jacobi--Anger identity $\int_0^{2\pi}\mathrm d\phi\,e^{\im u\cos\phi}=2\pi J_0(u)$, with $u=\kappa|\mathbf x_{\mathrm t}|$. The substitutions $s=\beta_{\mathrm p}(\kappa)$ and $s=\varkappa(\kappa)$ in the respective radial integrals remove the square-root denominators at grazing incidence and yield the following standard formulas~\cite{Arnoldus_Foley_2002,morse1953methods}.

\begin{proposition}[Regular one-dimensional Green-kernel integrals]
\label{prop:nonsingular_G}
Let $\mathbf x_{\mathrm t}\in\hat{x}_3^\perp$ and $z\ne0$. With $J_0$ denoting the Bessel function of the first kind of order zero, the kernels in~\eqref{eq:G_pr}--\eqref{eq:G_ev} have the equivalent representations
\begin{align}
G_{\mathrm{pr}}(\mathbf x_{\mathrm t}+z\hat{x}_3)
&=\frac{\im}{4\pi}\int_0^k\mathrm ds\,
J_0\!\left(|\mathbf x_{\mathrm t}|\sqrt{k^2-s^2}\right)e^{\im s|z|},
\label{eq:G_pr_nonsing}\\
G_{\mathrm{ev}}(\mathbf x_{\mathrm t}+z\hat{x}_3)
&=\frac{1}{4\pi}\int_0^\infty\mathrm ds\,
J_0\!\left(|\mathbf x_{\mathrm t}|\sqrt{k^2+s^2}\right)e^{-s|z|}.
\label{eq:G_ev_nonsing}
\end{align}
The evanescent integral is absolutely convergent for $z\ne0$; on $z=0$ the corresponding extension for $|\mathbf x_{\mathrm t}|>0$ is interpreted as an Abel limit.
\end{proposition}
For distributed sources, these kernels enter the convolution~\eqref{eq:rfd_rotated_source_field_integrals}; setting $\gamma=\mathrm I_3$ recovers the fixed-frame formulation.

\section{Analytical Properties and Regularity of the Propagation Potential}
\label{sec:rfd_analytical_structure}

The regularity condition $\mathcal V_J\in C^{1,1}_{\mathrm{loc}}(U)$ in Proposition~\ref{prop:rfd_local_wellposedness} can be established from the source distribution $J$. Write $P_J(q)\coloneqq\varphi_{\mathrm{pr}}(q)$, $F_J(\mathbf x)\coloneqq\varphi(\mathbf x)$ and $E_J(q)\coloneqq\varphi_{\mathrm{ev}}(q)=F_J(\mathbf x)-P_J(q)$. The function spaces used below are $C_c^\infty(\mathbb R^3)$ (smooth compactly supported functions), $C_c^{0,\alpha}(\mathbb R^3)$ (compactly supported H\"older-continuous functions of exponent $0<\alpha<1$), and $L^\infty(D)$ (essentially bounded functions on a measurable set $D$). On an open manifold domain, $C^{1,1}_{\mathrm{loc}}$ denotes continuously differentiable functions with locally Lipschitz first derivatives, and $W^{2,\infty}_{\mathrm{loc}}$ denotes functions with locally essentially bounded weak derivatives through order two in coordinate charts~\cite{evans2010pde,olver2014pde}. The first proposition uses the smoothing effect of convolution with a smooth source; the second accounts explicitly for the moving splitting plane and establishes the weaker regularity sufficient for local well-posedness.

\begin{proposition}[Regularity of the rotated modal potential]
\label{prop:rfd_smooth_rotated_convolutions}
Let $J\in C_c^\infty(\mathbb R^3)$. Then $\varphi_{\mathrm{pr}},\varphi_{\mathrm{ev}}\in C^\infty(\mathcal M)$, the regular domain $\operatorname{Dom}(\mathcal V_J)$ is open, and $\eta,\mathcal V_J\in C^\infty(\operatorname{Dom}(\mathcal V_J))$.
\end{proposition}
\begin{proof}
The partial Weyl kernels of Appendix~\ref{sec:rfd_weyl_fixed_frame} are locally integrable: Eq.~\eqref{eq:G_pr_nonsing} gives a continuous propagating kernel with $|G_{\mathrm{pr}}|\leq k/(4\pi)$, and $G_{\mathrm{ev}}=G_k^+-G_{\mathrm{pr}}$ has a locally integrable $1/(4\pi|\mathbf s|)$ singularity at the origin. For either label $\iota\in\{\mathrm{pr},\mathrm{ev}\}$, changing variables in Eq.~\eqref{eq:rfd_rotated_source_field_integrals} gives the distributional pairing
\begin{equation}
\label{eq:rfd_rotated_distributional_pairing}
 \varphi_\iota(\mathbf x,\gamma)
 =\big\langle G_\iota(\mathbf s),J(\mathbf x-\gamma\mathbf s)\big\rangle_{\mathbf s}.
\end{equation}
On every compact configuration neighborhood, the test functions $\mathbf s\mapsto J(\mathbf x-\gamma\mathbf s)$ and their parameter derivatives have a common compact support. All spatial and rotational derivatives can therefore be transferred to the smooth source factor in this pairing with the fixed locally integrable kernel. Differentiation under the integral is justified at every order, including for configurations whose splitting plane meets the source support. Both partial fields and their squared amplitudes $p$ and $e$ are consequently smooth on $\mathcal M$. Continuity of $p$ makes $\operatorname{Dom}(\mathcal V_J)=\{p>0\}$ open, and the quotient $\eta=e/p=-\mathcal V_J$ is smooth on this domain.
\end{proof}

The differential in Eq.~\eqref{eq:rfd_propagation_potential_differential} is obtained with $\mathrm dp=2\operatorname{Re}(\varphi_{\mathrm{pr}}^*\mathrm d\varphi_{\mathrm{pr}})$ and $\mathrm de=2\operatorname{Re}(\varphi_{\mathrm{ev}}^*\mathrm d\varphi_{\mathrm{ev}})$. Zeros of the total field with $p>0$ lie within the regular domain, whereas $p\to0$ with $e$ tending to a positive value gives $\mathcal V_J\to-\infty$, producing the energetic barrier of Proposition~\ref{prop:rfd_regular_domain_barrier}. At simultaneous zeros the quotient depends on the relative vanishing orders of the partial fields. For local solvability, the smooth-source hypothesis can be relaxed to the conditions in the following proposition. The key observation is that the longitudinal cusp of the finite-spectrum propagating kernel generates hypersurface terms in its second configuration derivatives; integration against a distributed source controls these terms uniformly.

\begin{proposition}[Regularity for distributed sources]
\label{prop:rfd_weak_source_regularity}
Let $D\subset\mathbb R^3$ be bounded and measurable and let $J\in L^\infty(D)$ be extended by zero. Then $P_J\in C^{1,1}_{\mathrm{loc}}(\mathcal M)$. If $J\in C_c^{0,\alpha}(\mathbb R^3)$ for some $0<\alpha<1$, then $F_J\in C^{2,\alpha}_{\mathrm{loc}}(\mathbb R^3)$ and $E_J\in C^{1,1}_{\mathrm{loc}}(\mathcal M)$, so that $\mathcal V_J\in C^{1,1}_{\mathrm{loc}}(\operatorname{Dom}(\mathcal V_J))$. For general bounded sources as above, the same regularity of $\mathcal V_J$ holds on the open set of regular configurations with $\mathbf x\notin\operatorname{supp}J$.
\end{proposition}
\begin{proof}
Choose a relatively compact configuration chart and a compact neighborhood $C_0$ within it, and enclose the source support in a bounded ball. All estimates are local and uniform on $C_0$. With $\hat\alpha=\gamma\hat{x}_3$ and $\mathbf d=\mathbf x-\mathbf y$, set $s=\hat\alpha\cdot\mathbf d$ and $u=|\mathbf d|^2-s^2$. By Eq.~\eqref{eq:rfd_rotated_propagating_kernel}, the propagating kernel is $K_{\mathrm{pr}}(\mathbf d,\gamma)=\mathscr K(u,|s|)$, where
\begin{equation}
\label{eq:rfd_weak_regular_kernel}
\mathscr K(u,v)=\frac{\im k}{4\pi}\int_0^1\!\mathrm dt\,J_0\!\left(k\sqrt{u}\sqrt{1-t^2}\right)e^{\im kvt}.
\end{equation}
The entire power series of the Bessel factor in $u$, together with the compact spectral interval, makes $\mathscr K$ smooth in $(u,v)$ with bounded derivatives on compact sets, including at $u=0$ and $v=0$. The functions $u,s$ and their configuration derivatives through order two are bounded uniformly on $C_0$ and the source-support ball. The nonsmooth dependence is confined to $|s|$, for which the distributional Hessian in chart coordinates $q^a,q^b$ is
\begin{equation}
\label{eq:rfd_moving_plane_second_derivative}
\partial_a\partial_b|s|=\operatorname{sgn}(s)\,\partial_a\partial_b s+2\delta(s)\,\partial_a s\,\partial_b s.
\end{equation}
Here $s=0$ is a regular hypersurface in configuration space for each fixed $\mathbf y$, since the derivative with respect to $\mathbf x$ is $\hat\alpha\ne0$. The distributional chain rule then expresses each second configuration derivative of $K_{\mathrm{pr}}$ as a bounded bulk term and the hypersurface term $2\partial_v\mathscr K(u,0)\delta(s)\partial_a s\,\partial_b s$. The latter includes the motion of the splitting plane induced by orientation derivatives.

To establish the bounds for an $L^\infty$ source without requiring its pointwise trace on a plane, take smooth compactly supported mollifications $J_j$ of the zero extension of $J$, with common bounded support, $\|J_j\|_{L^\infty(\mathbb R^3)}\leq\|J\|_{L^\infty(D)}$ and $J_j\to J$ in $L^1(\mathbb R^3)$. For these sources the preceding distributional derivatives can be integrated against $J_j$. The coarea formula in $\mathbf y$ converts the hypersurface term to an integral over the plane $\hat\alpha\cdot(\mathbf x-\mathbf y)=0$, since $|\nabla_{\mathbf y}s|=1$. Its intersection with the common source-support ball has uniformly bounded area. Thus the bulk terms are bounded by a constant times $\|J_j\|_{L^1(\mathbb R^3)}$, and the plane terms by a constant times $\|J_j\|_{L^\infty(\mathbb R^3)}$. The kernel and its first configuration derivatives are locally bounded as well, giving uniform $W^{2,\infty}$ bounds for the smooth convolutions $P_{J_j}$ on compact coordinate neighborhoods. Boundedness of the kernel and convergence of the sources in $L^1(\mathbb R^3)$ yield uniform convergence $P_{J_j}\to P_J$ there. Weak-star compactness of the weak derivatives through order two, identified by distributional convergence, gives $P_J\in W^{2,\infty}_{\mathrm{loc}}(\mathcal M)$. The continuous convolution agrees with the locally $C^{1,1}$ representative of this Sobolev class. Finally, local elliptic Schauder regularity for $(\nabla^2+k^2)F_J=-J$ gives $F_J\in C^{2,\alpha}_{\mathrm{loc}}(\mathbb R^3)$ for H\"older-continuous sources. For bounded sources, $F_J$ is smooth on source-free observation neighborhoods by interior elliptic regularity. In both cases $E_J=F_J-P_J$ has locally Lipschitz first derivatives on the stated region, and the quotient $\mathcal V_J=-|E_J|^2/|P_J|^2$ has the same regularity wherever $P_J\ne0$. These are precisely the potential-regularity conditions needed in Proposition~\ref{prop:rfd_local_wellposedness}.
\end{proof}

\section{Spectral-Axis Dependence of the Propagation Potential}
\label{app:rfd_point_source_rotation}

The inclusion of the local spectral frame in the extended configuration space (Section~\ref{sec:rotated_weyl_field_decomposition}) is motivated by the dependence of the spectral partition on the direction along which the radiation field is resolved. We establish this dependence analytically for a point source and investigate its angular and spatial structure numerically, before considering an asymmetric coherent source with both polar and azimuthal variation. These results provide the spectral basis for the directional properties of propagaton motion discussed in Section~\ref{sec:rfd_interpretation}. For a unit point source at the origin, $G_k^+(\mathbf x)=e^{\im kr}/(4\pi r)$, $r=|\mathbf x|>0$, and $k=2\pi/\lambda$. With $\hat r=\mathbf x/r$ and $\hat\alpha=\gamma\hat{x}_3$, let $\theta=\arccos|\hat\alpha\cdot\hat r|\in[0,\pi/2]$. The partial kernels have units of inverse length, while $r/\lambda$, $4\pi rG_\iota$ ($\iota=\mathrm{pr},\mathrm{ev}$), and $\mathcal V_G$ are dimensionless; angles are plotted in degrees. Expressing Eq.~\eqref{eq:rfd_rotated_propagating_kernel} in the dimensionless integration variable $t$ gives

\begin{equation}
\label{eq:rfd_point_pr_finite}
G_{\mathrm{pr}}(r,\theta)=\frac{\im k}{4\pi}\int_0^1\!\mathrm dt\,J_0\!\left(kr\sin\theta\sqrt{1-t^2}\right)e^{\im kr\cos\theta\,t}.
\end{equation}
The complementary kernel is $G_{\mathrm{ev}}=G_k^+-G_{\mathrm{pr}}$, independently evaluable off the splitting plane through Eq.~\eqref{eq:rfd_rotated_evanescent_kernel}. The dependence of both kernels on the orientation of the spectral frame is captured by $\theta$.

\begin{proposition}[Nonconstant propagation potential for an outgoing point source]
\label{prop:rfd_point_orientation_contrast}
At a fixed observation point with $r=\lambda/4$, the propagation potential $\mathcal V_G=-|G_{\mathrm{ev}}|^2/|G_{\mathrm{pr}}|^2$ takes the distinct values $-1/2$ and $0$ when the splitting axis is respectively parallel and perpendicular to the source--observation direction. The potential therefore distinguishes spectral-axis orientations even for a spherically symmetric total field.
\end{proposition}
\begin{proof}
For $\theta=0$, the Bessel factor in~\eqref{eq:rfd_point_pr_finite} equals one; for $\theta=\pi/2$, use $\int_0^1\!\mathrm dt\,J_0(a\sqrt{1-t^2})=\sin a/a$. Direct integration gives
\begin{align}
G_{\mathrm{pr}}(r,0)&=\frac{e^{\im kr}-1}{4\pi r}, &G_{\mathrm{ev}}(r,0)&=\frac{1}{4\pi r},\\
G_{\mathrm{pr}}(r,\pi/2)&=\frac{\im\sin(kr)}{4\pi r}, &G_{\mathrm{ev}}(r,\pi/2)&=\frac{\cos(kr)}{4\pi r}.
\label{eq:rfd_point_special_components}
\end{align}
Where the denominators are nonzero, these yield $\mathcal V_G(r,0)=-[4\sin^2(kr/2)]^{-1}$ and $\mathcal V_G(r,\pi/2)=-\cot^2(kr)$. At $kr=\pi/2$ both propagating components are nonzero and their potentials are $-1/2$ and $0$, respectively, as claimed.
\end{proof}

The axial expression $G_{\mathrm{ev}}(r,0)=1/(4\pi r)$ exhibits algebraic radial decay despite the exponential attenuation of individual evanescent modes. The continuous spectrum includes longitudinal attenuation constants arbitrarily close to zero, so that integration of the near-grazing sector produces the $1/r$ term. The corresponding relative strength $-\mathcal V_G(r,0)=[4\sin^2(kr/2)]^{-1}$, wherever defined, remains appreciable at arbitrarily large distances. The far-field limit therefore retains a nonuniform dependence on the chosen spectral axis, a property of the coherent amplitude partition rather than of the canonical field energy (Appendix~\ref{app:pure_scalar}).

\begin{remark}[Reality of the evanescent kernel]
\label{rem:rfd_real_evanescent}
The real Bessel and attenuation factors in Eq.~\eqref{eq:rfd_rotated_evanescent_kernel} give a real kernel $G_{\mathrm{ev}}$, including its boundary value away from the source singularity. For a uniform-phase source $J=e^{\im\vartheta_0}J_{\mathrm r}$ with $J_{\mathrm r}$ real, $\varphi_{\mathrm{ev}}=e^{\im\vartheta_0}\varphi_{\mathrm{ev,r}}$ with $\varphi_{\mathrm{ev,r}}$ real. On the regular domain, wherever zero is a regular value of this differentiable real function, its zero set is a hypersurface on which $\mathcal V=0$, with transverse quadratic behavior $\mathcal V\simeq-c\,\varphi_{\mathrm{ev,r}}^2$ ($c>0$). For a differentiable complex partial field with zero a regular value as a map into $\mathbb R^2$, the zero set instead has codimension two. For the point source, the real normalized amplitude $4\pi rG_{\mathrm{ev}}$ vanishes on curves in the $(r,\theta)$ plane; the branch containing the quarter-wavelength configuration of Proposition~\ref{prop:rfd_point_orientation_contrast} passes approximately through $(r/\lambda,\theta)=(0.25,90^\circ)$, $(0.30,73.9^\circ)$, $(0.40,54.9^\circ)$, $(0.50,43.8^\circ)$ and $(1.00,21.9^\circ)$. At regular differentiable zeros of $\varphi_{\mathrm{ev}}$, $\operatorname{grad}_g\mathcal V=0$; if the gradient is locally Lipschitz, uniqueness also precludes a nonstationary massless trajectory from reaching such an equilibrium in finite $\tau$. In contrast, a conservative massive propagaton of positive configuration-space energy attains its maximum energetically permitted geometrical speed upon reaching a regular zero, by Eq.~\eqref{eq:rfd_configuration_energy_identity} (Section~\ref{sec:rfd_interpretation}). For singular point sources the differentiability hypotheses are to be checked locally, particularly at the splitting plane; Appendix~\ref{sec:rfd_analytical_structure} establishes regularity for distributed sources.

\end{remark}

\begin{figure}[!tbp]
\centering\includegraphics[width=0.90\linewidth]{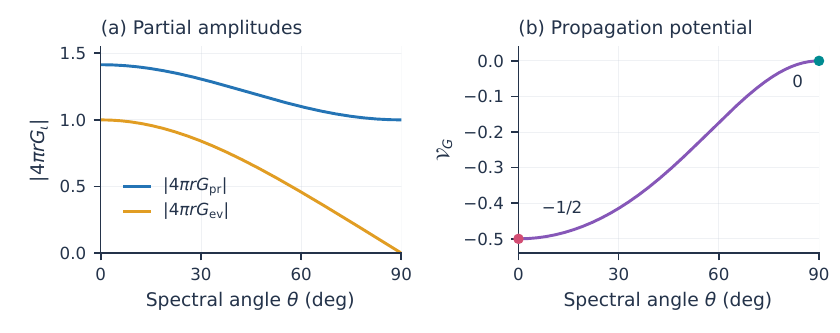}
\caption{Point-source angular dependence at $r/\lambda=1/4$. (a) Dimensionless partial magnitudes $|4\pi rG_\iota|$; (b) propagation potential $\mathcal V_G$. The exact endpoint values follow from Proposition~\ref{prop:rfd_point_orientation_contrast}.}
\label{fig:rfd_point_angular}
\end{figure}

At $r/\lambda=1/4$, Fig.~\ref{fig:rfd_point_angular} resolves the exact contrast of Proposition~\ref{prop:rfd_point_orientation_contrast} into a continuous angular profile. The decrease of the evanescent contribution toward zero is accompanied by an increase of the potential toward its upper bound, demonstrating how a change of spectral axis at a fixed spatial point supplies a driving direction in the rotational sector of $\mathcal M$.

\begin{figure}[!tbp]
\centering\includegraphics[width=0.72\linewidth]{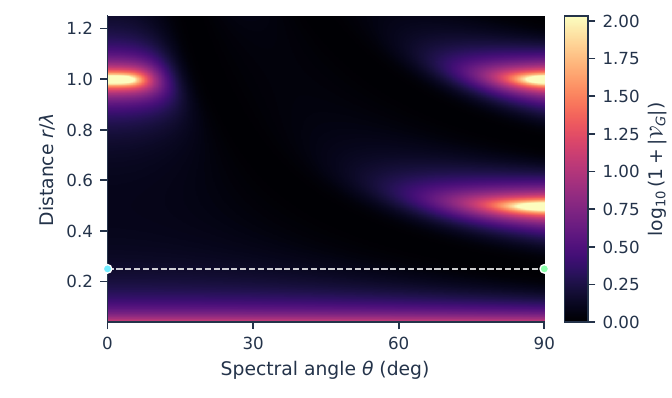}
\caption{Orientation--distance structure of the point-source potential. Color represents $\log_{10}(1+|\mathcal V_G|)$, with values above the displayed range shown at its upper limit. The dashed line marks $r/\lambda=1/4$.}
\label{fig:rfd_point_heatmap}
\end{figure}

Figure~\ref{fig:rfd_point_heatmap} extends the angular analysis to $0.04\le r/\lambda\le1.25$. The low-potential-magnitude bands trace regions of relative propagating purity, whereas large values of $\log_{10}(1+|\mathcal V_G|)$ can arise near small propagating amplitudes, the source of the energetic barriers discussed in Proposition~\ref{prop:rfd_regular_domain_barrier}. Their coexistence illustrates how spectral geometry can organize both equilibrium configurations and restrictions on accessible motion even in the smooth total field away from the point source. For the $128$-node Gauss--Legendre evaluation of Eq.~\eqref{eq:rfd_point_pr_finite}, doubling the order at $36$ test configurations changes $4\pi rG_{\mathrm{pr}}$ by at most $3.3\times10^{-14}$; independent improper-integral evaluations of $4\pi rG_{\mathrm{ev}}$ at six off-plane configurations agree within $5.0\times10^{-14}$.

\begin{figure}[!t]
\centering\includegraphics[width=0.94\linewidth]{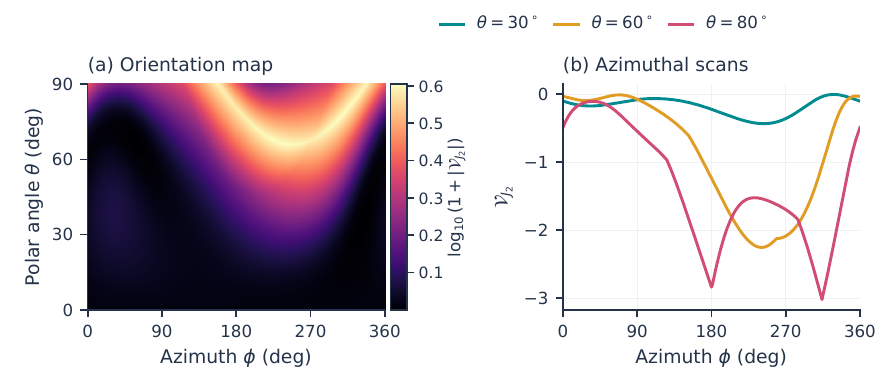}
\caption{Orientation dependence for the asymmetric coherent source~\eqref{eq:rfd_two_source_definition} at fixed $\mathbf x_0$. (a) $\log_{10}(1+|\mathcal V_{J_2}|)$; (b) potential along three fixed polar angles. Angles are expressed in degrees; the potential and its logarithmic transform are dimensionless.}
\label{fig:rfd_two_source_orientation}
\end{figure}

For the isotropic Green function, a change of spectral axis at fixed observation point is equivalent at the kernel level to a rotation of the separation vector with fixed spectral axis. To examine the orientation structure for a source with a distinguished spatial geometry and complex relative phase, consider two coherent point contributions,

\begin{equation}
\label{eq:rfd_two_source_definition}
J_2(\mathbf y)=\delta(\mathbf y-\mathbf y_1)+a_2\delta(\mathbf y-\mathbf y_2),\qquad
\mathbf y_{1,2}/\lambda=(\mp0.16,0,0),\quad a_2=0.68e^{\im\pi/3}.
\end{equation}
At $\mathbf x_0/\lambda=(0.24,0.19,0.46)$, let $\hat\alpha(\theta,\phi)=(\sin\theta\cos\phi,\sin\theta\sin\phi,\cos\theta)$. The propagating field is the weighted sum of Eq.~\eqref{eq:rfd_point_pr_finite} evaluated at $r_j=|\mathbf d_j|$ and $\arccos(|\hat\alpha\cdot\mathbf d_j|/r_j)$, with $\mathbf d_j=\mathbf x_0-\mathbf y_j$; subtraction from the total field gives the evanescent component. Axis reversal permits representation on the hemisphere $0\le\theta\le90^\circ$, $0\le\phi\le360^\circ$, with antipodal identification on its boundary (Section~\ref{sec:rfd_axial_momentum_reduction}). The potential in Fig.~\ref{fig:rfd_two_source_orientation} varies with both polar angle and azimuth: at $\theta=60^\circ$, for $\phi=0^\circ,90^\circ,180^\circ,270^\circ$ it takes the respective values $-0.02975$, $-0.07171$, $-1.23669$ and $-2.08515$, with fixed $|\varphi_{J_2}(\mathbf x_0)|=0.2307959\,\lambda^{-1}$. The minimum sampled $|\varphi_{\mathrm{pr},J_2}|=0.09672\,\lambda^{-1}$ on the $181\times361$ grid shows that this substantial contrast occurs with a finite propagating denominator throughout the sampled orientations. It reflects coherent interference of the partial responses from the two sources and supplies a genuinely two-angle potential on the position--axis space. At six selected configurations, quadrature-order doubling changes the propagating field by at most $1.8\times10^{-15}\,\lambda^{-1}$, and independent evanescent integration agrees within $2.3\times10^{-13}\,\lambda^{-1}$.

The corners in the azimuthal scans of Fig.~\ref{fig:rfd_two_source_orientation} arise when the signed longitudinal separation of either source crosses zero, producing the cusp in $|\hat\alpha\cdot\mathbf d_j|$. For distributed sources the convolution regularizes this structure under the hypotheses of Propositions~\ref{prop:rfd_smooth_rotated_convolutions} and~\ref{prop:rfd_weak_source_regularity}. Moreover, a smooth unit-mass approximate identity $J_\varepsilon\in C_c^\infty$ centered at the origin gives partial fields converging to the corresponding kernels at the fixed nonsingular observation point and the two orientations of Proposition~\ref{prop:rfd_point_orientation_contrast}. Since both limiting propagating amplitudes are nonzero and the limiting potentials differ by $1/2$, continuity of the quotient preserves an orientation contrast for sufficiently localized sources. The same argument applies to the asymmetric pair of Eq.~\eqref{eq:rfd_two_source_definition}, approximated by two smooth localized bumps with the specified complex weights. Thus the nontrivial angular structure persists for classes of distributed sources whose regularity, by Appendix~\ref{sec:rfd_analytical_structure}, supports locally well-posed propagaton dynamics under the metric hypotheses of Proposition~\ref{prop:rfd_local_wellposedness}.

\FloatBarrier

\section{Quotient Geometry and the Metric Symmetry Condition}
\label{app:rfd_quotient_technical}

To establish Proposition~\ref{prop:rfd_five_dimensional_reduction}, let $\mathcal U=\operatorname{Dom}(\mathcal V)$ and use the free, proper right action of the compact group $\mathcal G_{\mathrm{ax}}$ on $\mathcal U$. The projection~\eqref{eq:rfd_quotient_projection} identifies $\mathcal Q=\mathcal U/\mathcal G_{\mathrm{ax}}$ with an open subset of $\mathbb R^3\times S^2$. For an axially invariant metric $g$, the horizontal spaces $\mathcal H_q=(\ker(\mathrm d\mathfrak p)_q)^{\perp_g}$ define the quotient metric by Riemannian submersion~\cite{LeeRiemannian2018,MarsdenRatiu1999}: for $\bar q=\mathfrak p(q)$ and $\bar u,\bar v\in T_{\bar q}\mathcal Q$, with horizontal lifts $u^{\mathrm h},v^{\mathrm h}\in\mathcal H_q$,
\begin{equation}
\label{eq:rfd_quotient_metric_definition}
\bar g_{\bar q}(\bar u,\bar v)
=g_q(u^{\mathrm h},v^{\mathrm h}),\qquad
(\mathrm d\mathfrak p)_q u^{\mathrm h}=\bar u,\quad
(\mathrm d\mathfrak p)_q v^{\mathrm h}=\bar v.
\end{equation}
The isometric group action makes this definition independent of the representative $q$. For the product metric $g_0$ of Section~\ref{sec:rfd_geometric_dynamics}, horizontality is equivalent to $\boldsymbol\omega\cdot\hat{x}_3=0$, where $\gamma^{-1}\dot\gamma=\mathrm A(\boldsymbol\omega)$. Since $\mathrm d\hat\alpha/\mathrm d\tau=\gamma(\boldsymbol\omega\times\hat{x}_3)$, the horizontal rotational speed satisfies $|\mathrm d\hat\alpha/\mathrm d\tau|=|\boldsymbol\omega|$, giving $\bar g=\mathrm d\mathbf x\cdot\mathrm d\mathbf x+\ell^2\mathrm d\hat\alpha\cdot\mathrm d\hat\alpha$, as used in Section~\ref{sec:rfd_interpretation}. By Corollary~\ref{prop:rfd_potential_differential_symmetry}, $\mathcal V=\bar{\mathcal V}\circ\mathfrak p$ for a unique quotient scalar. Its gradient is horizontal, and for horizontal $w$ metric duality gives $g(\operatorname{grad}_g\mathcal V,w)=\mathrm d\bar{\mathcal V}[(\mathrm d\mathfrak p)_q w]$, hence $(\mathrm d\mathfrak p)_q(\operatorname{grad}_g\mathcal V)=\operatorname{grad}_{\bar g}\bar{\mathcal V}$. This establishes the massless reduction. For $m_*>0$, the condition $\mathcal J_{\mathrm{ax}}=m_*g(\dot q,Z)=0$ is equivalent to horizontality and is preserved by~\eqref{eq:rfd_axial_noether_integral}. For any horizontal curve $c$, the Killing identity gives $g(\nabla^g_{\dot c}\dot c,Z)=\frac{\mathrm d}{\mathrm d\tau}g(\dot c,Z)-g(\dot c,\nabla^g_{\dot c}Z)=0$. Its acceleration is therefore horizontal and projects to the Levi--Civita acceleration on $\mathcal Q$ by the submersion property. Projection of the unified law gives the reduced inertial equation, including relaxation, in Proposition~\ref{prop:rfd_five_dimensional_reduction}. Conversely, the horizontal lift of a reduced solution through a specified initial frame satisfies the full equation by these same identities; uniqueness follows from Proposition~\ref{prop:rfd_local_wellposedness} under its regularity hypotheses.

At nonzero axial momentum, reduction generally introduces additional connection- and momentum-dependent terms. The kinetic action is invariant under a fixed axial rotation $h\in\mathcal G_{\mathrm{ax}}$; an evolution-dependent rotation $h(\tau)=h(\theta(\tau))$ instead gives the squared speed $\|\dot q\|_g^2+2\dot\theta g(\dot q,Z)+\dot\theta^2\|Z\|_g^2$, by invariance of the metric. Metric compatibility is essential even for first-order equivariance (Proposition~\ref{prop:rfd_gradient_equivariance}): in a local chart with axial angle $\psi$ and coordinates $y,z^1,\ldots,z^4$, take the axially invariant scalar $\mathcal W=-y^2$, $\xi(\psi)=\tfrac12\sin\psi$, and the positive metric
\begin{equation}
\label{eq:rfd_noninvariant_metric_example}
g=(\mathrm dy)^2+2\xi(\psi)\,\mathrm dy\,\mathrm d\psi
 + (\mathrm d\psi)^2+\sum_{j=1}^4(\mathrm dz^j)^2.
\end{equation}
Inverting its $y$--$\psi$ block gives
\begin{equation}
\label{eq:rfd_noninvariant_gradient_example}
\operatorname{grad}_g\mathcal W
=-\frac{2y}{1-\xi(\psi)^2}(\partial_y-\xi(\psi)\partial_\psi).
\end{equation}
The gradient varies along the axial orbit despite the invariance of $\mathcal W$, illustrating how the geometry of $\mathcal U$ determines whether the spectral symmetry extends to the dynamics.

\section{Numerical Integration of Point-Source Propagatons}
\label{app:rfd_numerical_dynamics}

The point-source potential and its derivatives are evaluated from the finite spectral integral~\eqref{eq:rfd_point_pr_finite}, with $G_{\mathrm{ev}}=G_k^+-G_{\mathrm{pr}}$. The quotient metric of Appendix~\ref{app:rfd_quotient_technical} restricts to $g=\mathrm dx^2+\mathrm dy^2+\ell^2\mathrm d\psi^2$ on the planar sector $\mathbf x=(x,y,0)$, $\hat\alpha=(\cos\psi,\sin\psi,0)$. Indeed, the reflection $\sigma$ through the plane $x_3=0$ acts on the position--axis space by $(\mathbf x,\hat\alpha)\mapsto(\sigma\mathbf x,\sigma\hat\alpha)$, preserving both the quotient product metric and the point-source potential, which depends on $|\mathbf x|$ and $|\hat\alpha\cdot\mathbf x|$. Its fixed-point set is totally geodesic, and the gradient of the invariant potential is tangent to it. Both reduced dynamical laws (Proposition~\ref{prop:rfd_five_dimensional_reduction}) therefore preserve the sector for initial velocities tangent to it. For the distributional point source, the hypotheses of Proposition~\ref{prop:rfd_local_wellposedness} hold on the regular open region away from the source and splitting-plane cusp, where the potential is smooth. The computed trajectories remain within this region, with $\hat\alpha\cdot\hat r\geq0.339$ at the sampled configurations and nonvanishing propagating denominators.

With the normalized coordinates $Q=(X,Y,\psi)$, $s=\tau/T_0$ and $\rho=\ell/\lambda$ of Section~\ref{sec:rfd_interpretation}, the constant metric is $\widetilde g=\operatorname{diag}(1,1,\rho^2)$ in the planar chart. The massless equation becomes $Q'=\bar\zeta\widetilde g^{-1}\partial_Q\mathcal V$, and the conservative massive equation is $\bar m Q''=\widetilde g^{-1}\partial_Q\mathcal V$, with $\partial_Q\mathcal V$ denoting the column of coordinate derivatives. The geometrical parameters, normalized dynamical coefficients and initial conditions are given in Table~\ref{tab:rfd_parameters}. The units are $[\lambda]=L$, $[m_*]=M$, $[\Lambda]=ML^2/T_0^2$, $[\nu]=M/T_0$, $[\zeta]=L^2/T_0$, with $T_0$ a reference evolution scale. The potential, normalized energies, $s$, spatial coordinates and $\mu$ are dimensionless; $\psi$ and $\vartheta_{\mathrm F}$ are reported in degrees. The implementation uses $128$-node Gauss--Legendre quadrature with analytical first derivatives and an adaptive eighth-order explicit Runge--Kutta method (DOP853)~\cite{rfdHairer1993} at relative tolerance $10^{-10}$, absolute tolerance $10^{-12}$ and maximum step $0.002$ in $s$. Independent centered finite differences give a maximum derivative discrepancy of $2.22\times10^{-10}$ at the test configuration. Quadrature refinements with $64$, $96$ and $192$ nodes yield endpoint differences below $10^{-14}$ in the Euclidean norm of the normalized chart coordinates; loose-to-strict integration refinements give endpoint differences below $4\times10^{-13}$. The trajectories are sampled at $651$ equally spaced evolution parameters, and angular diagnostics are evaluated for dimensionless projected speeds exceeding $0.01$.

\begin{figure}[!t]
\centering
\includegraphics[width=0.98\linewidth]{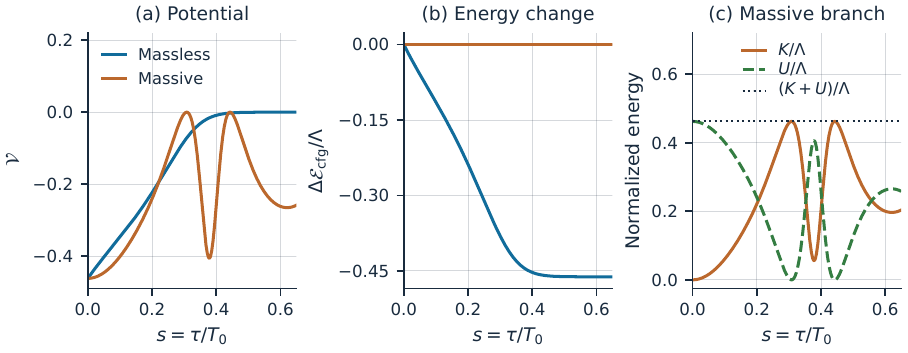}
\caption{Energetic structure of point-source propagaton dynamics. (a) Propagation potential for both branches; (b) normalized change in configuration-space energy, $\Delta\mathcal E_{\mathrm{cfg}}/\Lambda=(\mathcal E_{\mathrm{cfg}}(s)-\mathcal E_{\mathrm{cfg}}(0))/\Lambda$; (c) kinetic energy $K$, potential energy $U=-\Lambda\mathcal V$ and their sum for the conservative massive branch, normalized by $\Lambda$. All plotted quantities and the evolution parameter $s$ are dimensionless.}
\label{fig:rfd_pointsource_numerical_diagnostics}
\end{figure}

The energetic structure of the trajectories follows from Eq.~\eqref{eq:rfd_configuration_energy_identity}. Figure~\ref{fig:rfd_pointsource_numerical_diagnostics}(a,b) shows the evolution of $\mathcal V$ and the corresponding change in configuration-space energy for both branches. In the massive branch, the normalized kinetic energy is $K/\Lambda=(\bar m/2)(X'^2+Y'^2+\rho^2\psi'^2)$, and the normalized potential energy is $U/\Lambda=-\mathcal V$. Their complementary variation in Fig.~\ref{fig:rfd_pointsource_numerical_diagnostics}(c) illustrates the energy exchange discussed in Section~\ref{sec:rfd_interpretation}: near the two spectral-purity maxima the potential energy approaches zero and the kinetic energy approaches the conserved total. The sampled massive energy has a peak-to-peak variation of $6.05\times10^{-15}$ after normalization by $\Lambda$, while the massless potential increases monotonically. The zero axial momentum of the massive propagaton follows from its initial rest state and the horizontal lift of the reduced solution (Propositions~\ref{prop:rfd_axial_noether} and~\ref{prop:rfd_five_dimensional_reduction}). The planar solver also tests the distinct Noether momentum associated with simultaneous rotations of position and spectral axis about the plane normal, $\bar m(XY'-YX'+\rho^2\psi')$, whose maximum absolute sampled value is $1.25\times10^{-16}$. This provides a separate check of the rotational symmetry of the point-source problem.

\FloatBarrier

\begingroup
\footnotesize
\setlength{\bibsep}{0pt}
\bibliographystyle{unsrtnat}

\endgroup

\end{document}